\documentclass[aps,prx,superscriptaddress,twocolumn,longbibliography,reprint]{revtex4-2}

\pdfoutput=1

\usepackage[T1]{fontenc}
\usepackage{graphicx}
\usepackage{amsmath,amsfonts,amssymb,mathtools}
\usepackage[dvipsnames]{xcolor}
\usepackage{subfigure}
\usepackage{physics}
\usepackage{url}
\usepackage{booktabs}
\usepackage{hyperref}

\definecolor{emerald}{rgb}{0.31, 0.78, 0.47}
\definecolor{blue(ncs)}{rgb}{0.0, 0.53, 0.74}

\hypersetup{
     colorlinks=true,
     linkcolor=blue(ncs),
     filecolor=blue,
     citecolor=emerald,      
     urlcolor =blue(ncs),
}
	    
\definecolor{lightgray}{rgb}{0.9,0.9,0.9}	    
\definecolor{green}{rgb}{0,0.5,0}
\definecolor{red}{rgb}{1,0,0}
\definecolor{blue}{rgb}{0,0,0.5}

\long\def\symbolfootnote[#1]#2{\begingroup%
\def\thefootnote{\fnsymbol{footnote}}\footnotetext[#1]{#2}\footnotemark[#1]\endgroup}

\DeclareMathAlphabet{\pazocal}{OMS}{zplm}{m}{n}

\newcommand{\bo}[1]{\boldsymbol{#1}}
\newcommand{\D}{\mathrm{d}}

\begin{document}
\raggedbottom

\title{Sidewall Metrology Protocol for Superconducting Qubit Devices}

\author{J{\o}rgen Garn{\ae}s$^\ddag$}
\email{jg@dfm.dk}
\altaffiliation{$^\ddag$Contributed equally to this work}
\affiliation{Danish Fundamental Metrology, 2970 H{\o}rsholm, Denmark}

\author{Henrik S.~Guttesen$^\ddag$}
\affiliation{Danish Fundamental Metrology, 2970 H{\o}rsholm, Denmark}

\author{Astrid T.~R{\o}mer$^\ddag$}
\affiliation{Danish Fundamental Metrology, 2970 H{\o}rsholm, Denmark}

\author{Oscar Danielsson}
\affiliation{Danish Fundamental Metrology, 2970 H{\o}rsholm, Denmark}

\author{Emily P.~Duggan}
  \affiliation{Danish Fundamental Metrology, 2970 H{\o}rsholm, Denmark}

\author{Zhenhai Sun}
  \affiliation{NNF Quantum Computing Programme, Niels Bohr Institute, University of Copenhagen, Denmark}
  
\author{Asen Lyubenov Georgiev}
  \affiliation{Physikalisch-Technische Bundesanstalt, Bundesallee 100, 38116, Braunschweig, Germany}
  
  \author{Li Wei Chang}
  \affiliation{Physikalisch-Technische Bundesanstalt, Bundesallee 100, 38116, Braunschweig, Germany}

\author{Morten Kjaergaard}
  \affiliation{NNF Quantum Computing Programme, Niels Bohr Institute, University of Copenhagen, Denmark}

\author{Mark Bieler}
  \affiliation{Physikalisch-Technische Bundesanstalt, Bundesallee 100, 38116, Braunschweig, Germany}

\author{Lukas Grünhaupt}
  \affiliation{Physikalisch-Technische Bundesanstalt, Bundesallee 100, 38116, Braunschweig, Germany}

\author{Sabbir A.~Khan}
\email{skh@dfm.dk}

\affiliation{Danish Fundamental Metrology, 2970 H{\o}rsholm, Denmark}  

\date{\today}

\begin{abstract}
Sidewall morphology is a potentially important factor influencing superconducting qubit performance, yet its quantitative characterization remains challenging due to the absence of standardized measurement protocols. In this work, we develop a scanning probe microscopy-based measurement framework for the quantitative characterization of superconducting qubit sidewalls. The framework introduces a refined root mean square lateral sidewall roughness descriptor and a new curvature-equivalent radius for quantifying vertical sidewall morphology. The descriptors and measurement procedures are applied to different components of qubit devices, forming the basis of a proposed sidewall characterization protocol. This provides a foundation for quantitative comparisons of fabrication processes across laboratories, enabling future investigations of correlations between sidewall geometry and qubit performance.
\end{abstract}

\maketitle

\begingroup
\renewcommand{\thefootnote}{*}
\footnotetext{These authors contributed equally to this work.}
\endgroup

%
%%
%%%
\section{Introduction}
%%%
%%
%
Superconducting qubits are considered one of the leading technological platforms for building scalable quantum computers~\cite{KjaergaardEA20, Ezratty23}. However, achieving reliable and consistent device performance remains a major challenge~\cite{McRaeEA21, Siddiqi21, BurnettEA19}. Even for devices with the same design and layout, qubit performance often varies both between fabrication batches and within the same chip~\cite{VanDammeEA24, KimEA26}. Structural properties related to materials and fabrication processes are believed to be important factors influencing the consistency of superconducting qubit performance~\cite{Siddiqi21, Martinis22, StavengaEA23, KhanEA20}. Recent studies have highlighted sidewall profiles and angles as structural parameters that may influence the coherence times of superconducting qubit devices~\cite{QuintanaEA14, KimEA25, MurthyEA25}. Sidewall profiles and angles are usually investigated using cross-sectional focused ion beam–scanning electron microscopy (FIB-SEM) or high-resolution transmission electron microscopy (HR-TEM). Although these techniques provide high-resolution two-dimensional imaging, obtaining statistically representative quantitative information across devices remains challenging, making it difficult to establish statistically meaningful correlations with qubit performance~\cite{MurthyEA25, AlghadeerEA25, VerjauwEA22}. In order to draw reliable conclusions about the effect of sidewall properties on qubit lifetime, it is important to extract detailed quantitative information on the structural profile. Scanning probe microscopy (SPM) can potentially provide quantitative information on both sidewall angles and associated morphology with high spatial resolution. 

SPM-based metrology has been studied for characterizing semiconductor sidewall structures using tilted or three-dimensional atomic force microscopy (AFM) configurations~\cite{Kizu, MartinEA94, ChoEA11, KizuEA20, YooEA22}. However, the focus has primarily been on lateral line-edge roughness (LER), including its variation from the bottom to the top of the sidewall. Quantitative characterization of the full vertical sidewall profile, including local angles and morphological variations, has received comparatively less attention, particularly in the context of superconducting qubit devices, where such geometrical features may be relevant to loss mechanisms.

This work focuses on sidewall morphology and geometry, aiming to develop a framework for accurate, reproducible, and traceable measurements with associated quantitative descriptors. To establish these descriptors, we draw on a literature review and an analysis of relevant length scales and morphological features~\cite{LisenfeldEA19, KorshakovEA24, MurthyEA25, NPLEA25, BlandEA25, BhatiaEA2026, LuEA26}. The selected descriptors are intended to capture sidewall features that can be measured using AFM and relevant to device variability and loss mechanisms in superconducting qubits. We experimentally demonstrate the measurement and quantitative analysis of these descriptors on the sidewalls of key components of superconducting qubit devices. Based on these experimental results, we propose a sidewall measurement and data analysis protocol that can facilitate future statistical studies correlating sidewall features with qubit performance.

\begin{figure*}[t!bh]
	\centering
	\includegraphics[width=0.9\linewidth]{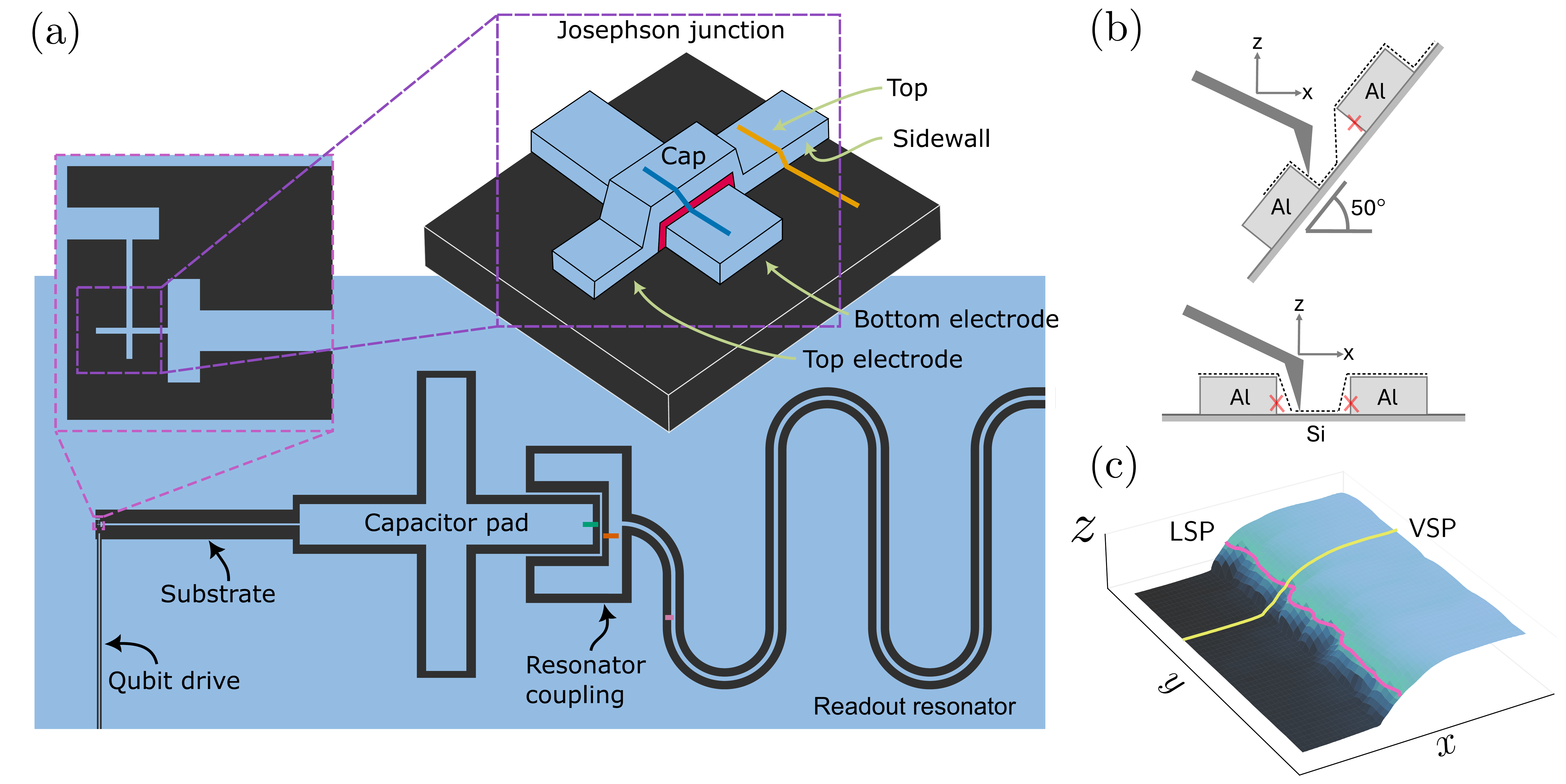}
    \caption{(a) Blueprint of Al-based (blue color is Al) superconducting qubit investigated in the present study (device number 1-7). Each qubit comprises a cross-shaped capacitor pad connected via an Al/AlO$_{x}$/Al Josephson junction to a surrounding Al layer. Each qubit is capacitively coupled (``resonator coupling'') to a readout resonator, and each readout resonator is inductively coupled to a feedline (not shown in the figure). A microwave driveline is capacitively coupled to the qubit island to enable controlled qubit driving ($\ket{0} \to \ket{1}$). The five colored lines (two in zoom inset and three in the main figure) show the positions of sidewall profiles shown in Figs.~\ref{fig:Correlations} and \ref{fig:VSPofCapandJJ}. (b) Schematic of the tilted-sample AFM approach enabling quantitative analysis of sidewall geometries, including undercut structures. The red crosses indicate vertical sidewall regions that are not accessible and cannot be reconstructed using the conventional non-tilted approach. (c) Schematic illustrating a Vertical Sidewall Profile (VSP) $Z_{\mathrm{VSP},i}(x)$  and a Lateral Sidewall Profile (LSP) $X_{\mathrm{LSP},i}(y)$. }
	\label{fig:Sketch}
\end{figure*}
%

%
%%
%%%
\section{Theoretical Considerations}
\label{sec:Theory}
%%%
%%
%

%
%%
\subsection{Transmon qubits, coherence, and resonators}
\label{sec:Recap}
Transmon qubits are defined by the two-state subspace consisting of the ground state $\ket{0}$ and the first excited state $\ket{1}$ of a circuit containing a Josephson junction connected to a capacitance (see Fig.~\ref{fig:Sketch} (a)) in the regime where $E_J / E_C \gg 1$, where $E_J$ is the Josephson energy and $E_C$ the charging energy~\cite{ManetiMotta}.

The qubit relaxation time ($T_1$) characterizes the duration over which an excited qubit relaxes back into the ground state, dissipating energy into the environment. For independent and Markovian decoherence processes $i$, relaxation times add reciprocally, $1/T_1 = \sum_i 1/T_1^i$, making any individual $T_1^i$ a lower bound of the total relaxation time. For transmon qubits, leading channels are commonly believed to be residual two-level systems at interfaces~\cite{LisenfeldEA19, Bland2EA26}, dielectric losses~\cite{EunEA23, SmirnovEA24}, Purcell loss through the readout resonator~\cite{Purcell1995, HouckEA08, BeaulieuEA26}, quasiparticle tunneling~\cite{JoyezEA94, deVisserEA11, KamenovEA23}, and radiative losses~\cite{BasharinEA26, DeDominicsEA26}. We note that the precise nature of the detrimental two-level systems is the topic of ongoing efforts to understand, but candidates include atomic tunneling defects, surface contaminants and oxides, adsorbates~\cite{LisenfeldEA19}, grain defects~\cite{NPLEA25}, and localized vibrons~\cite{TynerEA25}. For a broader, qualitative overview of sources limiting the performance of transmon qubits, we refer to Appendix \ref{sec:QubitCoherence}.

Sidewall morphology is also of importance for superconducting readout resonators, where dielectric and conductor losses may limit the attainable quality factor~\cite{LangEA26, BhatiaEA2026, LuEA26}. The etched sidewalls define mesoscopic metal–substrate and metal–vacuum interfaces that can exhibit enhanced electric-field participation and may constitute relevant sources of dissipation through two-level systems and other interfacial loss mechanisms. Moreover, the sidewall geometry may affect local field enhancement and the effective participation of surface oxides.

\subsection{Geometrical interfaces as sources of loss}
\label{sec:SourcesOfLoss}
The interfaces of a transmon qubit components (metal-air, metal-substrate, oxide-metal, etc.) each contribute to the dielectric loss relaxation time as 
\begin{equation}
T_1^{\mathrm{dielectric}} = \omega \sum_i p_i \tan \delta_i,
\label{eq:dielectricloss}
\end{equation}
where $\omega$ is the qubit frequency, and $p_i$ and $\tan \delta_i$ the participation ratio and loss tangent of interface $i$, respectively~\cite{Martinis22}. The participation ratio of the stored electrical energy in interface $i$ is given by 
\begin{equation}
  p_i = \frac{\epsilon_i}{2W} \int_A \D \bo{r}~t_i \lvert \bo{E}_i \rvert^2, 
  \label{eq:participation}
\end{equation}
where $\epsilon_i$ is the interface dielectric constant, $t_i$ the interface thickness, $\bo{E}_i$ the interface surface electric field, and $W$ the total capacitor energy. 

Gauss' law establishes that the surface charge density (which is directly proportional to the electric field) is inversely proportional to the radius of curvature for a surface protrusion, $\rho(\bo{r}_{\mathrm{prot}})$, i.e.,
\begin{equation}
    \lvert \bo{E}_i(\bo{r}_{\mathrm{prot}}) \rvert \propto \frac{1}{\rho(\bo{r}_{\mathrm{prot}})},
    \label{eq:electricfield}
\end{equation}
so dielectric losses may in general be expected to be more prominent at surface protrusions with small radii of curvature. The radius of curvature can be directly derived from the measured height profile. The surface topography probed by AFM may therefore be used to shed light on the likely impact of dielectric losses of the qubit. We caution, however, that for a correlation to be established in a statistically sound and meaningful fashion, a comprehensive understanding of the various decoherence channels and their relative contribution must first be established. We also caution that the qubits operate in a low charge density regime governed by superconducting electrodynamics and finite penetration depth (further commented on below) in which the entire geometry (such as capacitor pad spacing and orientation) may eventually decide whether the granularity and surface topography plays a significant or insignificant role in terms of dielectric losses. Recent studies have highlighted the importance and distribution of dielectric losses via parasitic two-level systems located at interfaces, which ultimately may put stringent conditions on surface purity and composition~\cite{BlandEA25, LisenfeldEA19, PurkayasthaEA25, KimEA25, MurthyEA25}.

In many quantum devices, AFM-resolved curvature on the scale of 10-100 nm (see also Sec.~\ref{sec:ProbeSpec}) are not merely descriptive of the surface morphology, but also fall in a range over which microwave currents and electromagnetic fields are redistributed. For example, a 200 nm Al film is expected to display a superconducting penetration depth around $\lambda = 60$ nm~\cite{LopezEA25} and a coherence length in vicinity to the range of $\xi \approx 100$-$250$ nm depending on the mean free path.

In addition to the planar surfaces, sidewalls constitute an important part of the metal–substrate–vacuum interface and may contribute disproportionately to dielectric losses~\cite{KimEA25, MurthyEA25}. Sidewall morphology, including its inclination, curvature, and roughness, determines both the local electric-field distribution and the interfacial area available for hosting parasitic two-level systems. This may be particularly important in Josephson junctions, in which the electrical field concentrates~\cite{ZengEA15, Rhee06, KorshakovEA24}. Variations in the sidewall profiles arising from the fabrication process may alter the participation of lossy dielectric regions, making quantitative characterization of the sidewall important with respect to the coherence of superconducting circuits~\cite{Siddiqi21, StavengaEA23}.

We emphasize that surface roughness should not be viewed as a direct measure of superconducting quality, but rather as a possible proxy for the microscopic film and interface properties that contribute to readout resonator loss. Rougher growth may accompany greater defect and grain-boundary densities, modified oxide/interface chemistry, increased effective interfacial area, or local thickness and field inhomogeneities, all of which can enhance dissipative channels such as two-level systems~\cite{HedrickEA26, KaruppannanEA25, AlmasouriEA26}. Electronic disorder itself is not necessarily detrimental. In disordered or granular superconductors electronic disorder can increase $T_c$ or kinetic inductance while preserving low losses~\cite{khan2023epitaxially, BardeenMattis58, GuptaEA25}. Roughness may serve as an experimentally accessible indicator of the underlying material state. No universal monotonic relation between surface roughness and internal quality factors should be expected across different materials and fabrication processes.

\subsection{Definition of relevant geometrical descriptors}
\label{sec:DefDescriptors}
In this subsection the theory for two sidewall morphology descriptors is laid out, one to characterize the lateral morphology and one to characterize the vertical (cross sectional) morphology. The theory behind the curvature-based probe radius estimation are also given. Additionally, the band pass filtering of the profiles is summarized briefly in Appendix~\ref{sec:DefGeomDescriptors}.

\subsubsection{Lateral sidewall profiles} 
The Lateral Sidewall Profile (LSP), $X_{\mathrm{LSP},i}(y)$, describes the lateral position of the sidewall surface projected onto a plane parallel to the substrate, where $y$ denotes the coordinate along the nominal sidewall direction, see Fig.~\ref{fig:Sketch} (b-c). The index $i$ identifies the individual profiles measured at different vertical positions along the sidewall from the foot of the sidewall to the top. Details of reference subtraction, the filtering procedure, and the practical descriptor calculation is provided in Appendix~\ref{sec:DefGeomDescriptors}.

The RMS lateral sidewall roughness, $R_q^{\mathrm{LSP}}$, is defined as the average RMS amplitude of the band pass filtered residual profiles:
\begin{equation}
R_q^{\mathrm{LSP}} = \frac{1}{N}
\sum_{i=1}^{N}
R_{q,i}^{\mathrm{LSP}},
\end{equation}
where $R_{q,i}^{\mathrm{LSP}}$ is the RMS amplitude calculated for the individual profile $i$, and $N$ is the number of measured profiles. 

\subsubsection{Vertical sidewall profiles and morphology} 
The Vertical Sidewall Profile (VSP), $Z_{\mathrm{VSP},j}(x)$, is defined as the cross section of the vertical sidewall, see Fig.~\ref{fig:Sketch} (c), where the index $j$ identifies the position along the y-direction of the vertical sidewall profile, and the y-direction is parallel to the nominal sidewall direction. A residual amplitude descriptor such as $R_q$ cannot be meaningfully defined as there is no meaningfully reference profile available. However, according to Eq.~(\ref{eq:electricfield}), dielectric losses may in general be expected to be more prominent at surface protrusions with small radii of curvature. Thus we propose the local vertical sidewall morphology to be characterized by a curvature-equivalent radius, $R_{\rho}$, deduced from the local radius of curvature of the vertical sidewall profiles, which weigh small radius of curvatures more strongly than large radius of curvatures. 

The curvature-equivalent radius, $R_{\rho}$, is defined as the inverse of the profile-averaged RMS curvature. The RMS-equivalent curvature is first determined for each individual profile and subsequently averaged over all measured profiles.

Small values of $R_{\rho}$ indicate the presence of numerous sharp surface features with small local radii of curvature, whereas larger values correspond to smoother sidewalls. The aucillary minimum radius of curvature, $\rho_{\min}$, is defined as the smallest local radius of curvature along the sidewall profile. The detailed definitions of the vertical sidewall profile $Z_{\mathrm{VSP},j}(x)$, filtering procedure, curvature calculation, RMS averaging, the curvature-equivalent radius and the minimum local radii of curvature are given in Appendix~\ref{sec:DefGeomDescriptors}.

\begin{figure*}[t!bh]
	\centering
	\includegraphics[width=\linewidth]{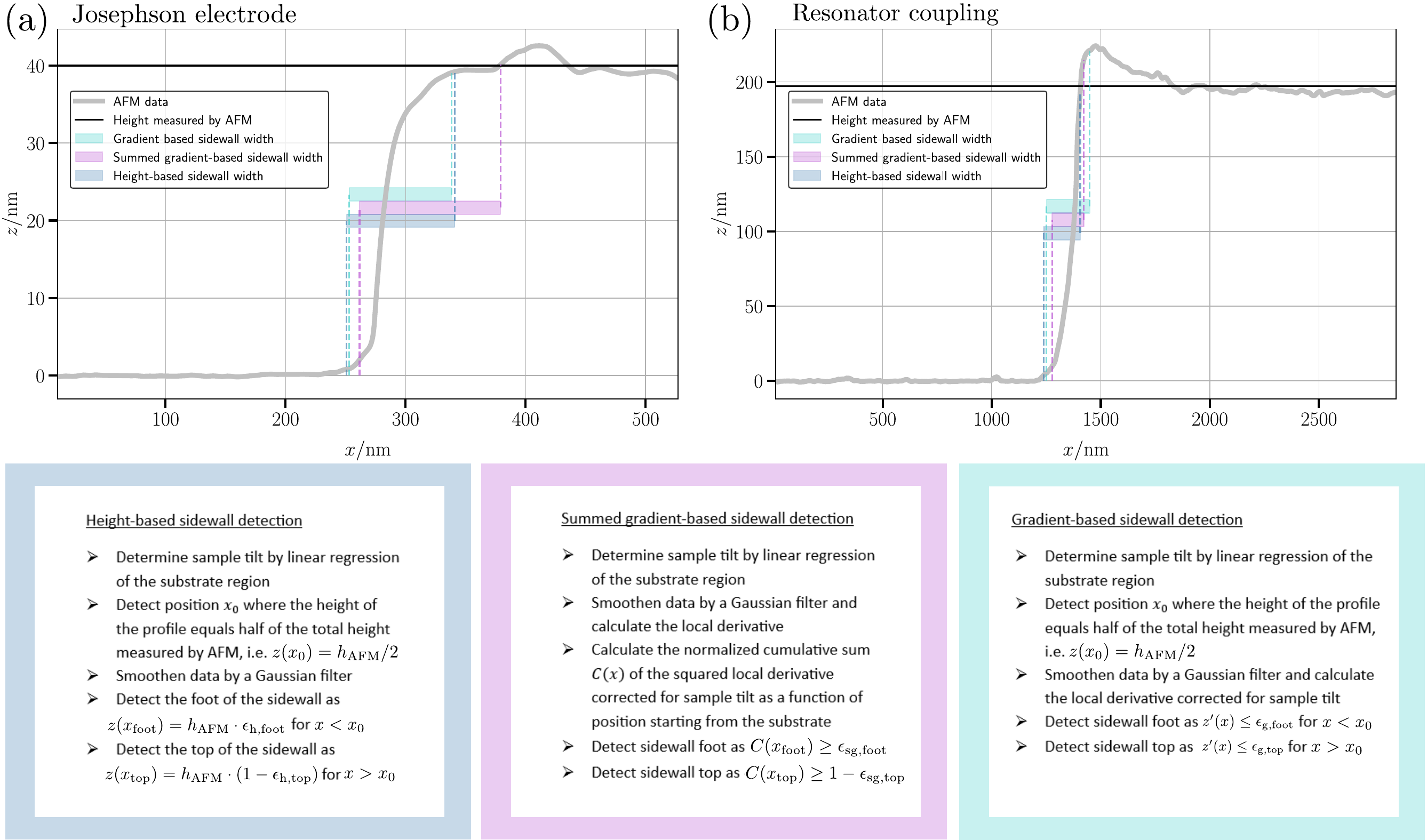}
    \caption{ (a) Example of a vertical sidewall profile of a Josephson junction electrode with indications of the detected sidewall region by use of the height-based method (shown in blue), summed gradient-based method (shown in magenta) and the gradient-based method (shown in turquoise). (b) Example of a vertical sidewall profile of a resonator coupling with indications of the detected sidewall region by use of the height-based method (shown in blue), summed gradient-based method (shown in magenta) and the gradient-based method (shown in turquoise). The three different methods are described in the text.}
    \label{fig:fig4}
\end{figure*}

\subsection{Curvature-Based Probe Radius Estimation}
In the measurements, the recorded surface profiles represent the surface topography convolved with the geometry of the AFM probe tip. Although geometrical deconvolution methods based on a known probe geometry exist, traceable reference standards for validating reconstructed probe geometries in the range $<$ 10~nm are, to the knowledge of the authors, not available. Therefore, we recommend that the descriptors (measurands) are specified together with the probe geometry used for the measurements, in analogy to the specification of stylus geometry in conventional roughness measurements. Thus, the AFM probe is regarded as an integral part of the measurement procedure. Surface features that cannot be resolved due to the probe geometry, such as hypothetical deep trenches with widths smaller than the probe diameter, are therefore not considered quantifiable with the present measurement approach. This approach is consistent with the selection of an appropriate S-filter prior to further analysis, which defines the lower spatial wavelength limit of the evaluated surface morphology.

In this work, an effective probe radius $\rho_{\mathrm{probe}}^{\mathrm{eff}}$ was estimated from the local curvature of multiple sharp protrusions on a tip characterization sample; details of the procedure are given in Appendix~\ref{sec:Curvature-BasedRadiusEstimation}.

\subsection{Algorithm for sidewall region detection}
One of the fundamental challenges in processing sidewall profiles is to develop a general method for identifying the sidewall region in the data. Depending on the structure of the sidewall and the way it was fabricated, this transition region can be more or less straight-forward to determine. In Fig.~\ref{fig:fig4}, we display three methods for sidewall step edge detection. The first method is based on knowledge of the step height and takes as input parameter the AFM-derived height, $h_{\mathrm{AFM}}$, obtained from a least-squares fit to the central third of the aluminium pad region. The foot of the sidewall is defined as the position where the profile displays a height of $z(x)=\epsilon_{h,\mathrm{foot}} h_{\mathrm{AFM}}$, where $\epsilon_{\mathrm{h},\mathrm{foot}}=0.02$ is used for the two examples shown in Fig.~\ref{fig:fig4}.  Similarly, the top is detected at the position where the profile displays a height of $z(x)=(1-\epsilon_{\mathrm{h},\mathrm{top}}) h_{\mathrm{AFM}}$, with  $\epsilon_{\mathrm{h},\mathrm{top}}=0.02$. The resulting sidewall region is displayed by the blue bars in Fig.~\ref{fig:fig4}, and labeled height-based sidewall width. To ensure a robust routine, the foot and top are detected starting from the sidewall midpoint, as defined by the position where $z(x)=h_{\mathrm{AFM}}/2$. In this way, we prevent roughness of the substrate or deposition structure to interfere with the step detection. 

In the second method, which is shown by the turquoise bars in Fig.~\ref{fig:fig4} and is denoted the gradient-based sidewall detection, the gradient is calculated along the sidewall profile, and the foot is determined as the position, where the local gradient becomes smaller than a a value, $\epsilon_{\mathrm{g},\mathrm{foot}}=0.05$, set in the analysis. Depending of the choice of this parameter, the detection sensitivity can be increased or decreased. Likewise, the top of the sidewall is determined as the position where local gradient becomes smaller than the set value $\epsilon_{\mathrm{g},\mathrm{top}}$. Similar to the first method, we detect foot and top starting from the middle of the sidewall.

In the third method, we instead use the cumulative sum of the squared gradient to detect the foot and top of the sidewall, starting the analysis from the substrate region. In Fig.~\ref{fig:fig4}, the result of this method is shown by the magenta bars, and the detection threshold is set by $\epsilon_{\mathrm{sg},\mathrm{foot}}$ and $\epsilon_{\mathrm{sg},\mathrm{top}}$ for the foot and top, respectively. In the present example we used $\epsilon_{\mathrm{sg},\mathrm{foot}} = 0.005$ and $\epsilon_{\mathrm{sg},\mathrm{top}} = 0.02$. 
The cumulative-gradient sidewall detection method defines the transition region based on the accumulated change in local sidewall angle. It therefore defines the transition from the distribution of geometrical changes along the profile rather than from an external height reference. This approach enables, e.g., protrusions or overshoots at the sidewall edge to be treated in a consistent manner. Since the method identifies the transition based on the accumulated gradient, the result is inherently influenced by the profile length over which the gradient is integrated, and it should be noted that, as opposed to the first two methods, the third method is more susceptible to roughness of the substrate, which can in principle lead to an overestimation of the sidewall width. 

Based on the observed profiles, including those shown in Fig.~\ref{fig:fig4}, the height-based sidewall detection method provides a robust and reproducible definition of the transition length. However, it requires an independent determination of the step height and is, by definition, not sensitive to the exact transition from the substrate to the sidewall or from the sidewall to the top plane of the pad, even when these are well defined. The gradient-based sidewall detection method also provides a robust and reproducible definition of the transition length, where the transition is defined as the region over which the local sidewall angle changes. Based on the observed profiles, the method is highly sensitive to a well defined transition between e.g. the substrate and the sidewall. Since the method identifies the transition from changes in the local sidewall angle, it is, by definition, independent of the height of the aluminum pad and by experience the transition point between the sidewall and the top plane is not always intuitively correct.

\begin{figure*}[t!bh]
	\centering
	\includegraphics[width=\linewidth]{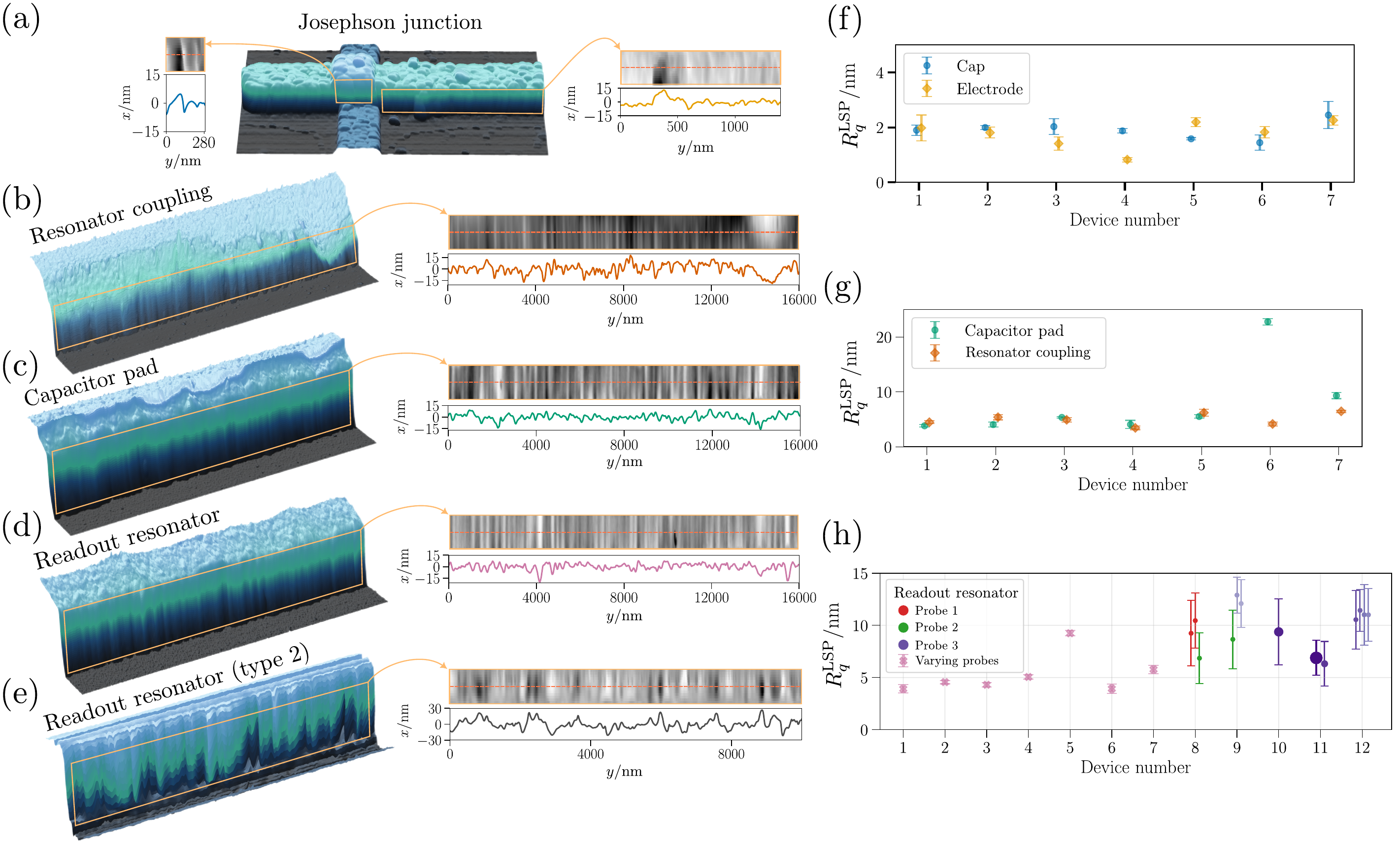}
    \caption{(a)-(e) AFM height maps of various circuit components and their sidewalls. Orange windows indicate the locations of the in-plane profile variations shown in grayscale images next to the maps. A lateral sidewall profile $X_{\mathrm{LSP},i}(y)$ corresponding to the middle cut of the orange windows are indicated below each grayscale map. (f) Average sidewall roughness calculated from multiple sidewall profiles acquired across seven wet-etched transmon qubit devices. (g) Same as (f) for linear circuit elements. (h) Profile-averaged sidewall roughness calculated for the readout resonators of the same seven devices as measured in (f) and (g), as well as five additional readout resonators from an alternative fabrication process, exemplified in (e) and referred to as type 2 (concerning devices 8-12). In (h) we provide information of the AFM probes used in the measurements to quantify the reproducibility of the results and dependence on the probe. For device 8, we observe that repeated measurements with Probe 1 as well as an additional measurement with Probe 2 give results that agree within the uncertainty. In the case of Probe 3, we invoke different shade and size of the data points to reflect the time that data was taken going from dark/large to bright/small for increasing times. The error bars represent the variation of the observed vertical sidewall profile $Z_{\mathrm{VSP},i}(x)$ extracted at different $z_{i}$-positions from the foot to the top of the aluminum pad. The relative difference in sidewall roughness between devices 11 and 12 is comparable to the small variation observed in repeated measurements using the same probe.}
	\label{fig:Correlations}
\end{figure*}

%
%%
%%%
\section{Results and discussion}
\label{sec:ResultsAndDiscussio}
%%%
%%
%

%
%%
\subsection{Lateral sidewall profiles}
Figure~\ref{fig:Correlations} summarizes the AFM characterization of the lateral sidewall morphology for the investigated superconducting qubits and readout resonators. The average lateral sidewall roughness, $R_q^{\mathrm{LSP}}$, ranges from approximately 2-5~nm for the Josephson junction structures to approximately 20~nm for the readout resonators. A sidewall with a visible local pit exhibits an increased roughness value of approximately 30~nm. The profile-to-profile variation of the extracted roughness values within the same image area, quantified by $\sigma\left(R_{q,i}^{\mathrm{LSP}}\right)$, ranges from approximately 0.3--0.6~nm for the qubit components and up to approximately 5~nm for the readout resonators. The extracted roughness values $R_q^{\mathrm{LSP}}$ and their associated variations demonstrate that the fabricated sidewalls exhibit measurable deviations from the ideal geometry, which can be quantified consistently using the proposed amplitude descriptors.

\begin{figure*}[t!bh]
	\centering
	\includegraphics[width=\linewidth]{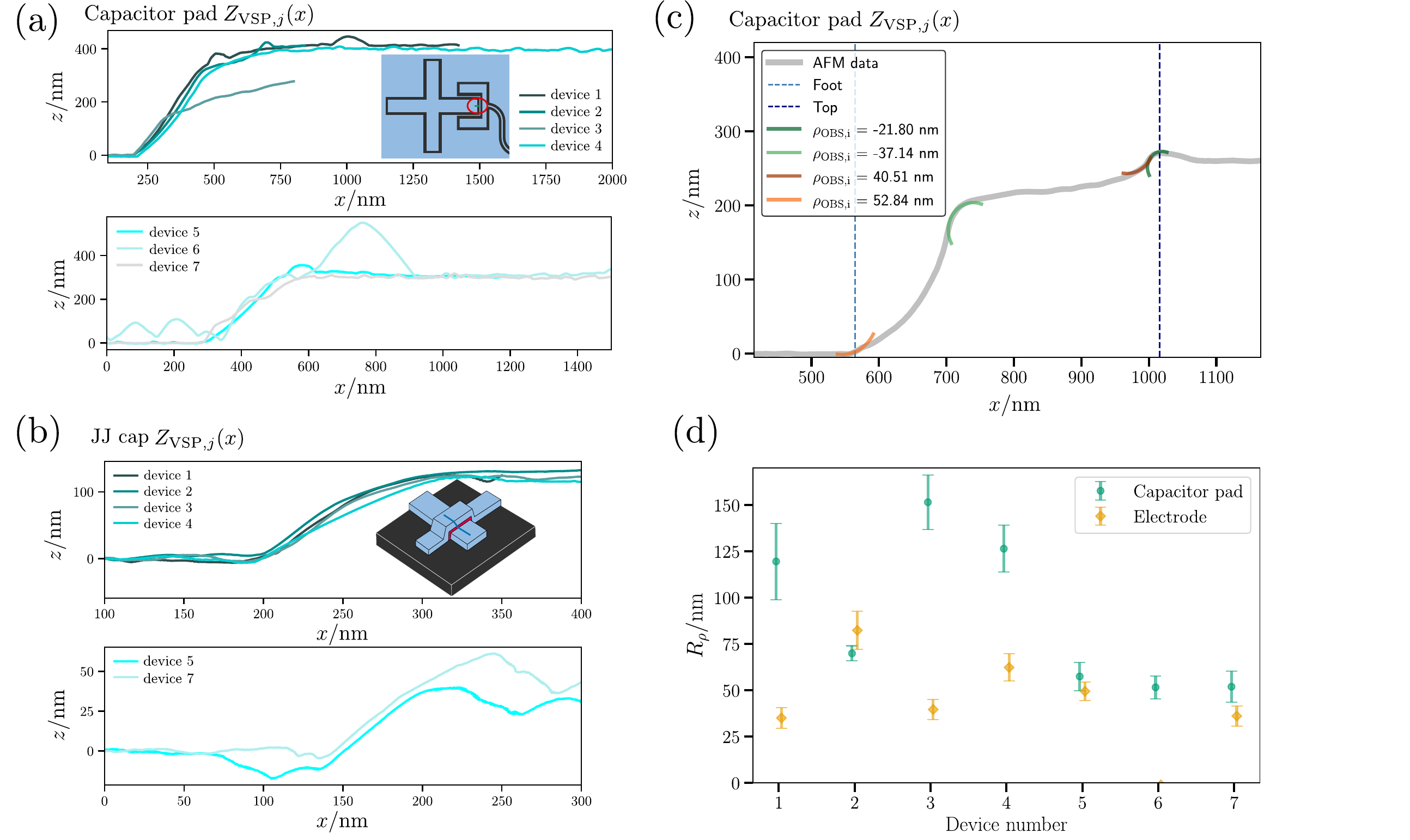}
    \caption{(a) Vertical sidewall profiles $Z_{\mathrm{VSP},i}^{f}(x)$ of the capacitor edge at the positions marked in the inset. (b) Similar profiles for the Josephson junction element ``Cap'' (the double-layer structure) at the positions marked in the inset. (c) Single profile of the capacitor edge on device 2. Circular arcs corresponding to the local radius of curvature $\rho$ are shown for the two smallest protrusions and two smallest depressions. (d) Effective radius of curvature $R_\rho=1/\overline{\kappa}_{\mathrm{RMS}}$ (see Eq.~\eqref{eq:kappaRMSaverage}) of the vertical profiles of JJ electrodes and capacitor pads as a function of device number.}
    \label{fig:VSPofCapandJJ}
\end{figure*}

\subsection{Vertical sidewall profiles}
Figure~\ref{fig:VSPofCapandJJ} summarizes the observed vertical sidewall profiles of a capacitor pad and the smaller Josephson junction (JJ) electrode. A qualitative assessment gives that most profiles on the same chip are similar, however, one profile (Capacitor pad of device 4) is significantly different, indicating that a single profile of one edge is not sufficient to make a general statement about the angle and shape of the sidewall on the chip. 

The single profile of the capacitor (device 2) in panel (c) shows a distinct sharp protrusion near the top of the aluminum pad, with an observed local minimum radius of $\rho_{\mathrm{OBS}} = -21.8~\mathrm{nm}$ (negative for protrusions), which is equal to the conservative estimate of the resolution limit of the probes. Corrected for the probe shape, the best estimate of the local radius of curvature is $\rho = \rho_{\mathrm{OBS}} - \rho_{\mathrm{probe}}^{\mathrm{eff}} = -21.8~\mathrm{nm} + 10.9~\mathrm{nm} = -10.9~\mathrm{nm}$. The associated measurement uncertainty from probe-to-probe variation is $6.3~\mathrm{nm}$, see Sec.~\ref{sec:ProbeSpec}. 

Panel (d) shows the profile-averaged RMS curvature $R_{\rho}$ of the capacitor pad and the JJ electrode for the different devices. The electrodes have values of $R_{\rho}$ in the range from 35 to 82 nm, while the capacitor pads have values in the range from 51 to 151 nm. The observed variation between sidewall profiles at the same spot ranges from 2 to 10 nm and is in general larger for larger $R_{\rho}$. A short-wavelength cutoff of $\lambda_{\mathrm{S}} = 8~\mathrm{nm}$ and $\lambda_{\mathrm{S}} = 25~\mathrm{nm}$ was applied to the capacitor and electrode profiles, respectively. The filtering strongly attenuates features with spatial wavelengths well below $8~\mathrm{nm}$ and $25~\mathrm{nm}$, respectively, and therefore limits the possible observation of radii of curvature well below these values.

\subsection{Sidewall angle distribution}
A way to visualize and analyze the sidewall geometry without compromising the sidewall sampling point density obtained in a tilted AFM scan is to bin the local sidewall angles into a (normalized) probability density function (PDF). A symmetry reference point for such an analysis is chosen to be the sidewall midpoint (defined above). For example, an idealized $S$-shaped sidewall with a single inflection point about which it permits symmetry, such as $z = h \arctan(x/w)$, would display identical sidewall angle PDFs on either side of the inflection point. We employ the height-based sidewall detection scheme to define the sidewall foot, top, and midpoint. 

\begin{figure*}[t!bh]
	\centering
	\includegraphics[width=\linewidth]{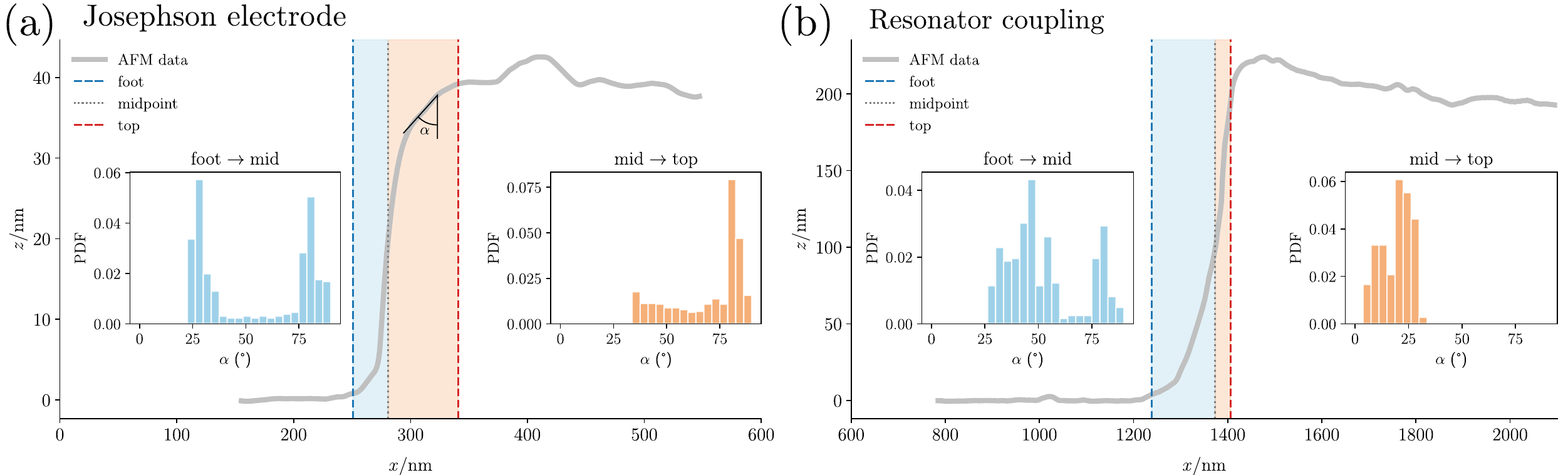}
    \caption{Sidewall angle analysis for a non-linear circuit element (device no.~5) (a) and a linear circuit element (device no.~6)) (b). In this analysis we identify the foot, midpoint, and top of the sidewall using the height-based sidewall scheme (cf.~Fig.~\ref{fig:fig4}) with $\epsilon_{\mathrm{h},\mathrm{foot}} = \epsilon_{\mathrm{h},\mathrm{top}} = 0.02$. Local sidewall angles $\alpha$, defined by the angle with the vertical, are binned to produce normalized probability distributions (PDF's) corresponding to the regions between the foot and midpoint and between the midpoint and the top, shown in the insets. The sidewall angle PDFs may be characterized by their median, $\bar{\alpha}$, and interquartile range, $\mathrm{IQR}$, resulting in: $\bar{\alpha}_{\mathrm{JJ},\mathrm{foot}\to\mathrm{mid}} =51^{\circ}$, $\mathrm{IQR}_{\mathrm{JJ},\mathrm{foot}\to\mathrm{mid}}=53^{\circ}$, $\bar{\alpha}_{\mathrm{JJ},\mathrm{mid}\to\mathrm{top}} =79^{\circ}$, $\mathrm{IQR}_{\mathrm{JJ},\mathrm{mid}\to\mathrm{top}}=27^{\circ}$ for (a), and $\bar{\alpha}_{\mathrm{Coupl},\mathrm{foot}\to\mathrm{mid}} =47^{\circ}$, $\mathrm{IQR}_{\mathrm{Coupl},\mathrm{foot}\to\mathrm{mid}}=28^{\circ}$, $\bar{\alpha}_{\mathrm{Coupl},\mathrm{mid}\to\mathrm{top}} =21^{\circ}$, $\mathrm{IQR}_{\mathrm{Coupl},\mathrm{mid}\to\mathrm{top}}=13^{\circ}$ for (b).}
    \label{fig:fig5}
\end{figure*}

Figure~\ref{fig:fig5} shows the resulting sidewall angle PDFs on either side of the sidewall midpoint for both a non-linear (a) and a linear circuit element (b). Both AFM profiles reveal partly shallow and wide sidewall angle PDFs, clear asymmetries about the midpoint, and some distributions even being bimodal~\footnote{The physical origin of a bimodal sidewall angle distribution may for example be a two-step taper, faceting, a change in etch mechanism partway up the wall, a re-entrant or bowed profile, or micro-trenching near the foot.}. These observations should be contrasted with the (more idealized) case of a straight edge with rounded corners, for which one usually expects a unimodal, sharply peaked PDF. The shallowness of the sidewalls (such as characterized by the median value $\bar{\alpha}_{\mathrm{JJ},\mathrm{foot}\to\mathrm{mid}} =51^{\circ}$ in Fig.~\ref{fig:fig5} (a)) may be expected to have consequences for the electrostatics of the device. Furthermore, the rather large widths of the sidewall angle distributions (such as characterized by the interquartile range $\mathrm{IQR}_{\mathrm{JJ},\mathrm{foot}\to\mathrm{mid}}=53^{\circ}$ in Fig.~\ref{fig:fig5} (a)) and their non-unimodality problematize simplified characterizations of the geometry, such as by using a single sidewall angle to describe its shape. Across devices and circuit features (even within a single qubit) a large variety of shapes and sidewall angle distributions are observed.

%
%%
%%%
\section{Recommended measurement protocol}
%%%
%%
%
The definitions of the sidewall descriptors and the analysis of the measurement results presented in this work form a practical, step-by-step measurement procedure. A summary of the sidewall descriptors and the essential elements defining the AFM measurement protocol are summarized in Table~\ref{tab:MesProt}.

\begin{table*}
\centering
\caption{Summary of sidewall descriptors and essential elements defining the AFM measurement protocol.}
\label{tab:MesProt}
\begin{tabular}{p{0.24\textwidth} p{0.22\textwidth} p{0.22\textwidth} p{0.28\textwidth}}
\toprule
\multicolumn{1}{c}{\textbf{Sidewall descriptors}} &
\multicolumn{1}{c}{\textbf{Probe specification}} &
\multicolumn{1}{c}{\textbf{Measurement parameters}} &
\multicolumn{1}{c}{\textbf{Data processing and analysis}} \\ \midrule
\begin{itemize}
\item $R^{\mathrm{LSP}}_q$: Lateral sidewall roughness, Sec.~\ref{sec:DefDescriptors}
\item $R_{\rho}$: Curvature equivalent radius, Sec.~\ref{sec:DefDescriptors}
\item $\rho_{\min}$: Minimum radius of curvature,  App.~\ref{sec:DefGeomDescriptors}
\item Sidewall angle $\alpha$ and width, Fig.~\ref{fig:fig4} 
\end{itemize}
&
\begin{itemize}
\item Probe specification is part of measurement, Sec.~\ref{sec:ProbeSpec}
\item Estimation of probe radius $\rho^{\mathbf{eff}}_{\mathrm{probe}}$, Eq.~\eqref{RhoEffEqualHalfOBS},
App.~\ref{sec:Curvature-BasedRadiusEstimation}
\item Uncertainty of probe radius $u(\rho^{\mathbf{eff}}_{\mathrm{probe}})$, Eq.~\eqref{uRhoEffProbe}
\end{itemize}
&
\begin{itemize}
\item Tilt of sample, scan length, scan direction, number of pixels, number of profiles analysed, Sec.~\ref{sec:InstrumentationMeasurement}
\item Validity criteria for measurements App.~\ref{sec:ValidityAndAcceptanceCriteria}
\item Mitigation strategies App.~\ref{Improving-Masurement-Repeatability}
\end{itemize}
&
\begin{itemize}
\item Leveling Sec.~\ref{sec:DataProcessing}
\item High- and low-pass filtering parameters $\lambda_s$, $\lambda_c$,  Sec.~\ref{sec:DefDescriptors}
\item Correction for probe shape Eq.~\eqref{CorrectedForTheProbe} and App.~\ref{TheoryCPRE-Approach}
\item Definitions of 
$R^{\mathrm{LSP}}_q$, Eq.~\eqref{eq:Rq_LSP}, 
$R_{\rho}$, Eq.~\eqref{eq:Rkappa},
$\rho_{\min}$, Eqs.~\eqref{eq:rhomin1}, \eqref{eq:rhomin2}
\item Sidewall width, Fig.~\ref{fig:fig4} and \ref{fig:fig5}
\item Evaluation of uncertainty 
\newline
Table~\ref{tab:variability}, App.~\ref{Improving-Masurement-Repeatability}
\item Limitations and precautions App.~\ref{Limitation} and~\ref{sec:ValidityAndAcceptanceCriteria}
\end{itemize}
\\ \bottomrule
\end{tabular}
\end{table*}
\begin{table*}
\centering
\caption{Standard uncertainty $u$ denotes the estimated uncertainty contribution associated with each source of variability for the measurands $R_q^{\mathrm{LSP}}$ and $R_\rho$. Residual variation $s$ denotes the experimentally observed standard deviation after mitigation. The corresponding quantities are denoted $u(R_q^{\mathrm{LSP}})$, $u(R_\rho)$, $s(R_q^{\mathrm{LSP}})$, and $s(R_\rho)$.}
\label{tab:variability}
\begin{tabular}{p{0.35\textwidth} p{0.13\textwidth} p{0.35\textwidth} p{0.12\textwidth}}
\toprule
\multicolumn{1}{c}{\textbf{Source of variability}} &
\multicolumn{1}{c}{\textbf{Standard }} &
\multicolumn{1}{c}{\textbf{Mitigation strategy}} &
\multicolumn{1}{c}{\textbf{Variation $s$}} \\
 & \multicolumn{1}{c}{\textbf{uncertainty $u$}} & & \multicolumn{1}{c}{\textbf{after}} \\ & & & \multicolumn{1}{c}{\textbf{mitigation}}
\\ \midrule
Probe-to-probe variation ($R^{\mathrm{LSP}}_q$) &
\multicolumn{1}{c}{10-20\%} &
Use of a single pre-characterized probe ($R^{\mathrm{LSP}}_q$) &
\multicolumn{1}{c}{$\leq$ 4\%} \\
Probe-to-probe variation ($R_\rho$) &
\multicolumn{1}{c}{5-8 nm} &
Use of a single pre-characterized probe ($R_\rho$) &
\multicolumn{1}{c}{3-5 nm} \\
Probe degradation and instability &
\multicolumn{1}{c}{5-15 \%} &
Probe stability monitoring &
\multicolumn{1}{c}{$\leq$ 4\%} \\
Spatial inhomogeneity of the sample &
\multicolumn{1}{c}{Varies} &
Averaging over large scan areas or multiple locations &
\multicolumn{1}{c}{Varies} \\
Particles and contamination & \multicolumn{1}{c}{$\lesssim$ 50\%} &
Particle removal and separate particle analysis &
\multicolumn{1}{c}{$<2\%$} \\
Additional experimental parameters (e.g. applied force) &
\multicolumn{1}{c}{$\lesssim$ 20\%} &
More detailed specification of measurement parameters &
\multicolumn{1}{c}{$\leq$ 4\%} \\
Data processing (outliers, algorithm implementations, sequence) &
\multicolumn{1}{c}{$\lesssim$ 10\%} &
Standardized data-processing procedure  &
\multicolumn{1}{c}{$\leq$ 4\%} \\ \bottomrule
\end{tabular}
\end{table*}

A key objective of a measurement protocol is to enable reproducible and comparable measurements across different laboratories and measurement systems. An integral part of such a protocol is therefore an assessment of the expected variability and the associated uncertainty of the reported sidewall descriptors. 

Table~\ref{tab:variability} lists the sources of variability and the associated standard uncertainty and include the influence of the probe used, possible contamination, experimental settings and data processing. For a detailed discussion of the source of variability and the associated standard deviation see Appendix~\ref{SourcesMeasurementVariability}.

\subsection{Improving measurement repeatability}

Repeatability describes the precision obtained when, for example, the same probe and measurement settings are used throughout a series of measurements. Under these conditions, relative differences in sidewall descriptors can typically be determined much more precisely than the variability observed when different probes or measurement parameters are used. Consequently, repeatability determines the sensitivity with which differences between sidewalls can be resolved over a short time span, whereas probe-to-probe variability and other effects primarily limit the comparability of measurements performed at different times or in different laboratories. 

Table~\ref{tab:variability} lists the mitigation strategies and the residual variation after mitigation. The residual variation $s(R_q^{\mathrm{LSP}})$ after the mitigation strategy has been applied is, in most cases, calculated from experimental observations. Similar residual variations are expected for $s(R_\rho)$; however, in contrast to $s(R_q^{\mathrm{LSP}})$, the values reported for $s(R_\rho)$ should be regarded as approximate estimates based on experimental observations and laboratory experience. For a detailed discussion of mitigation strategies for all sources of variation, see Appendix~\ref{Improving-Masurement-Repeatability}.

\subsection{Combined measurement uncertainty}
In conventional stylus-based roughness measurements the influence of the mechanical stylus used are usually not corrected for. In other words the measured roughness and associated combined measurement uncertainty are method depend. In this work we suggest a similar approach, that is, the measured result and associated uncertainty should be interpreted together with the shape of the AFM probe and data-processing used. Thus, a possible bias associated with the AFM approach using an AFM probe and the associated data-processing procedure is regarded as part of the measurement method and is therefore not treated as a separate uncertainty contribution.

In this work the combined measurement uncertainty include contributions associated with the use of different probes, probe degradation and instability, possible adsorbed particles and contamination, variations in measurement conditions, and data-processing procedures. In addition to these contributions, the influence of imperfect calibration of the microscope also contributes to the overall measurement uncertainty. Based on calibration measurements using certified reference standards, this contribution is estimated to be below 3\%. Based on the discussion and the variability estimates presented in Appendix~\ref{SourcesMeasurementVariability}, the combined measurement uncertainty ($k=1$) for the morphological descriptors investigated in this work is estimated to be in the range of 15-30\%.

%
%%
%%%
\section{Conclusion}
\label{sec:Conclusion}
%%%
%%
%
A protocol for superconducting qubit sidewalls characterization has been developed. The protocol introduces two morphological descriptors; a refinement of the RMS lateral sidewall roughness and a new curvature-equivalent radius for quantifying the vertical sidewall morphology, together with an auxiliary parameter describing the minimum radius of curvature along the profile. In addition, methods for defining the sidewall position and extracting the average sidewall angle have been established. The proposed descriptors and methods were applied to representative samples and shown to provide consistent and physically meaningful measures of sidewall geometry Measurement variability, uncertainty, and relevant limitations and precautions were also discussed to establish a systematic framework for sidewall characterization.The proposed protocol is expected to facilitate reproducible sidewall measurements and comparisons of fabrication processes across laboratories. Future work may extend the protocol through a comprehensive uncertainty analysis, additional morphological descriptors, and statistical correlations between sidewall characteristics and device performance.

%
%%
%%%
\section{Methods}
%%%
%%
%
%
%%
\subsection{Sample fabrication}
The transmon devices were fabricated on high-resistivity silicon substrates ($>20$ k$\Omega$cm). 200 nm-thick aluminum films were deposited on the substrate using a Plassys MEB550S E-beam evaporator. The qubit capacitors, readout resonators, feedlines, and control lines were patterned by E-beam lithography and subsequently wet-etched by the Transene Type D aluminum etchant. Josephson junctions were formed by Manhattan-style double-angle aluminum evaporation in the same Plassys evaporator, with an intermediate oxidation step. Electrical connections between the Josephson junctions and the remainder of the circuits were defined by e-beam lithography, followed by argon-ion milling and aluminum deposition in the Plassys evaporator. In addition, two chips containing a total of six aluminum readout resonators were fabricated separately, and are referred to as type 2 in Fig.~\ref{fig:Correlations}(e) and regards device number 8-12, shown in Fig.~\ref{fig:Correlations}(h). These resonators were fabricated on double polished C-plane sapphire $\mathrm{Al_2O_3}$. Wafers were pre-cleaned in acetone for 5 minutes in a sonic bath followed by oxygen (O$_2$) plasma cleaning for 5 minutes at a power of 200 W and an O$_2$ pressure of 1 mbar. For the base layer we deposit 100 nm of aluminium at a base pressure below 10$^{-7}$ mbar using an e-beam evaporator with a deposition rate of $0.5$ nm/s. 
To align subsequent lithography steps, we first pattern Ag markers using thermal evaporation and a lift-off process.  The resonators and feedline are wet-etched using a single PMMA layer developed using an IPA/water mixture (5:3) for 70 s and etched with ma-D 332 for approximately 4 minutes. The wafer is then profusely rinsed with water and spin dried and remaining resist stripped in 50$^\circ$C acetone for 10 minutes and a sonic bath for 3 minutes. Finally, we rinse the wafer with IPA and spin dry.

\subsection{Probe specification}
\label{sec:ProbeSpec}
In this work, AFM probe with a specified radius of curvature below 10~nm are used (Non-contact probes, spring constant $\sim45\,\mathrm{N/m}$ and with the probe located at the cantilever end (see Fig.~\ref{fig:Sketch} (b) and Ref.~\cite{ATEC_NC}). The effective radius of the batch of probes used - including wear - was estimated from images and profiles of a sample designed for AFM probe characterization and found to be
\begin{equation}
\rho_{\mathrm{probe}}^{\mathrm{eff}} = 10.9 \,\mathrm{nm}.
\end{equation}
and the associated standard uncertainty was conservatively estimated to be $u(\rho_{\mathrm{probe}}^{\mathrm{eff}}) = 6.3 \,\mathrm{nm}$. Details of the estimation, including representative profiles and images of the AFM probe-characterization sample, are given in Appendix~\ref{sec:ExperimantalCPRE}. 

\subsection{Instrumentation and measurement} 
\label{sec:InstrumentationMeasurement}
The tilted sidewalls were characterized using a high-resolution NX20 AFM (Park Systems, South Korea) with a scanning range of $100\,\mu\mathrm{m} \times 100\,\mu\mathrm{m} \times 15\,\mu\mathrm{m}$ in intermittent-contact mode. For optimal imaging of the sidewall morphology and geometrical shape, the scanning plane should ideally be parallel to the sidewall surface. However, the accessible sample tilt is limited by the AFM geometry and sample configuration, which prevent an exact alignment of the apriory unknown angle of the sidewalls. In this work, the samples were mounted on a $50^\circ$ wedge with the z direction of the AFM aligned roughly perpendicular to the sidewall, see Fig.~\ref{fig:Sketch} (b). 

The sidewalls were measured on seven test qubits distributed over two chips at locations, as indicated in Fig.~\ref{fig:Sketch}. In addition, sidewall profiles of five readout resonators were measured on the two chips containing readout resonators.

The scan length in the fast-scan direction was intended to include at least two long-wavelength cutoffs when possible. For the linear components (capacitor pads, coupling structures, and readout resonators), the long- and short-wavelength cutoff limits were selected as $\lambda_C = 8~\mu\mathrm{m}$ and $\lambda_S = 25~\mathrm{nm}$, respectively. The cutoff wavelength $\lambda_C$ and $\lambda_S$ were selected based on the physical considerations discussed in Sec.~\ref{sec:SourcesOfLoss} and taking the possible measurement range of the AFM into account also. Sampling per profile should be as high as possible, in this work 4096 pixels per lines. The number of profiles per area investigated are at least 32.

For the Josephson junction electrodes, the scan length was limited by the feature dimensions and was typically $2$--$3\,\mu\mathrm{m}$, maintaining a sampling interval well below the Nyquist criterion. The long- and short-wavelength cutoff limits were selected as $\lambda_C = 0.25~\mu\mathrm{m}$ and $\lambda_S = 25~\mathrm{nm}$ for the $R_q$ analysis and $\lambda_S = 8~\mathrm{nm}$ for the $R_{\rho}$ analysis. For the Josephson junction cap (that is, the 3D overlap structure), this length scale is comparable to the total width of the sidewall and no additional long-wavelength cutoff is applied. For the cap a short-wavelength cutoff $\lambda_S = 8\,\mathrm{nm}$ was chosen. Selected values are based on the physical considerations discussed in Sec.~\ref{sec:SourcesOfLoss}. Further considerations regarding measurement validity and acceptance criteria, particularly relevant to the reliable determination of the radius of curvature, are given in Appendix~\ref{sec:ValidityAndAcceptanceCriteria}

\subsection{Data Processing}
\label{sec:DataProcessing}
For general AFM data processing and calculation of lateral sidewall roughness, SPIP~\cite{SPIP} was used. A \texttt{python}-based analysis routine was developed for the calculation of the curvature-equivalent radius, sidewall width, and sidewall angle distribution.

For the lateral sidewall profiles, a least-squares linear reference profile was subtracted from each measured profile prior to filtering and calculation of the descriptors. All profiles were subsequently filtered using a Gaussian S-filter with a short-wavelength cutoff $\lambda_S$ and where appropriated an L-filter with a long-wavelength cutoff $\lambda_C$ described in Sec.~\ref{sec:InstrumentationMeasurement}, followed by calculation of the lateral roughness.

The profile-averaged RMS curvature, $\kappa_{\mathrm{RMS}}$, is independent of the rotation of the Cartesian coordinate system due to the use of the geometric curvature definition.

%
%%
%%%
\section{Acknowledgments}
%%%
%%
%
The author acknowledges the support of DFM, PTB, and NQCP facilities in carrying out this work. In some sections of the article, an AI tool was used for language and grammar corrections. SAK acknowledges funding from Danish Agency for Higher Education and Science, DIANA Quantum Center (DIANAQ), The Danish Chips Competence Center (DKCCC), DIRAC: A national CERN lighthouse initiative, Innovation Fund Denmark (IFD) projects: DeQD, QLIGHT, European Metrology projects: MetSuperQ (Project No. 23FUN08), NoQTeS (Project No. 23NRM04), DINAMO (Project No. 24DIT02). MK gratefully acknowledge support from the Novo Nordisk Foundation (grant no. NNF22SA0081175), the NNF Quantum Computing Programme (NQCP), Villum Foundation through a Villum Young Investigator grant (grant no. 37467), the Innovation Fund Denmark (grant no. 2081-00013B, DanQ), the U.S. Army Research Office (grant no.~W911NF-22-1-0042, NHyDTech), by the European Union through an ERC Starting Grant, (grant no.~101077479, NovADePro), and by the Carlsberg Foundation (grant no.~CF21-0343). Any opinions, findings, conclusions or recommendations expressed in this material are those of the author(s) and do not necessarily reflect the views of Army Research Office or the US Government. Views and opinions expressed are those of the author(s) only and do not necessarily reflect those of the European Union or the European Research Council. Neither the European Union nor the granting authority can be held responsible for them. MB and LG acknowledge funding by the German Federal Ministry of Research, Technology and Space (BMFTR) within the framework programme “Quantum technologies – from basic research to market” (contract number: 13N16158) and by the European Partnership on Metrology, co-financed from the European Union’s Horizon Europe Research and Innovation Programme and by the Participating States (contract number: 23FUN08 MetSuperQ). 

%
%%
%%%
\section{Author Contributions}
%%%
%%
%
S.A.K, J.G. conceived the idea, experimental plans. J.G, O.D and E.P.D performed the experiments. J.G, H.S.G, A.T.R. performed data analysis. Z.S., A.L.G., L.W.C. M.K., M.B. and L.G. provided transmon and resonator samples. J.G, H.S.G, A.T.R., S.A.K prepared the manuscript with the contribution from all other authors. S.A.K supervised the project.        

%
%%
%%%
\section{Competing financial interests}
%%%
%%
%
The authors declare no competing financial interests.

%
%%
%%%
\section{Data availability}
%%%
%%
%
The data supporting the findings in the study can be made available upon reasonable request to the corresponding authors.

%
%%
%%%
\section{Supplementary Information}
%%%
%%
%
Appendix A: Sources of qubit decoherence and dephasing; Appendix B: Definition of relevant geometrical descriptors; Appendix C: Curvature-Based Probe Radius Estimation (CPRE);
Appendix D: AFM measurement validity and acceptance criteria; Appendix E: Experimental estimation of the effective probe radius; Appendix F: Sources of measurement variability;
Appendix G: Improving measurement repeatability;
Appendix H: Relation between ISO Gaussian filtering and standard deviation formulation.

\bibliography{Refs}

\begin{appendix}
\onecolumngrid

\clearpage

%
%%
%%%
\section{Sources of qubit decoherence and dephasing}
\label{sec:QubitCoherence}
%%%
%%
%
Complementary to the qubit relaxation time, $T_1$, the qubit dephasing time ($T_{2}$) characterizes how long a superposition of the qubit’s basis states retains its relative phase before phase fluctuation in the environment cause the phase information to decay (even if the population of the states remains unchanged). It therefore measures the timescale over which quantum interference effects persist.

Table~\ref{tab:Keyfactors} provides a qualitative summary of key parameters that influence transmon qubit performance. Typical sources that limit coherence include dielectric loss, Purcell decay, quasiparticle tunneling, radiative loss, two-level systems, and infrared radiation. Typical important sources of dephasing include charge noise, flux noise, photon shot noise, and $1/f$ noise.

\begin{table}
\centering
\label{tab:Keyfactors}
\caption{Qualitative summary of key transmon qubit features and their qualitative impact on the relaxation time ($T_1$) and the dephasing time ($T_2$).}
\begin{tabular}{p{0.30\textwidth} p{0.35\textwidth} p{0.30\textwidth}}
\toprule
Parameter & Impact on $T_1$ & Impact on $T_2$ \\ 
\hline
Capacitor geometry & Interface and radiative loss & Charge noise sensitivity \\
Josephson junction size & Two level system and Josephson energy $E_J$ & Flux noise if looped \\
Resonator capacitive coupling & Purcell decay & Photon shot noise \\
Dielectric interfaces & Dielectric losses and heat dissipation & Two-level system induced frequency noise \\
Chip packaging & Radiative and IR losses & Stabilization of thermal environment \\
Filtering and shielding & Cut stray radiation, reduce quasiparticle generation, reduce radiative losses & Reducing dephasing via environment \\ \bottomrule
\end{tabular}
\end{table}
%

%
%%
%%%
\section{Definition of relevant geometrical descriptors}
\label{sec:DefGeomDescriptors}
%%%
%%
\subsection{Profile filtering} 
All profiles analyzed are first corrected by subtraction of a reference line if applicable. Then a short-wavelength cutoff filter is applied with the wavelength limit $\lambda_{S}$. This operation retain all spatial wave lengths longer than $\lambda_{S}$ while suppressing shorter wavelengths. Thus this will suppress noise and possible influence of the detailed probe structure on a length scale below $\lambda_{S}$. Where applicable also a long-wavelength cutoff filter is applied with the wavelength limit $\lambda_{C}$. This filter retain spatial wave lengths shorter than $\lambda_{C}$ while suppressing longer wave lengths. The intention is to remove the waviness of the surface morphology and only include the spatial bandwidth possibly relevant for the variation or losses. This selection of bandwidth is in analogy to roughness measurements on a longer scale.

\subsection{The lateral sidewall profile}
\begin{figure}[t!bh]
	\centering
	\includegraphics[width=0.3\linewidth]{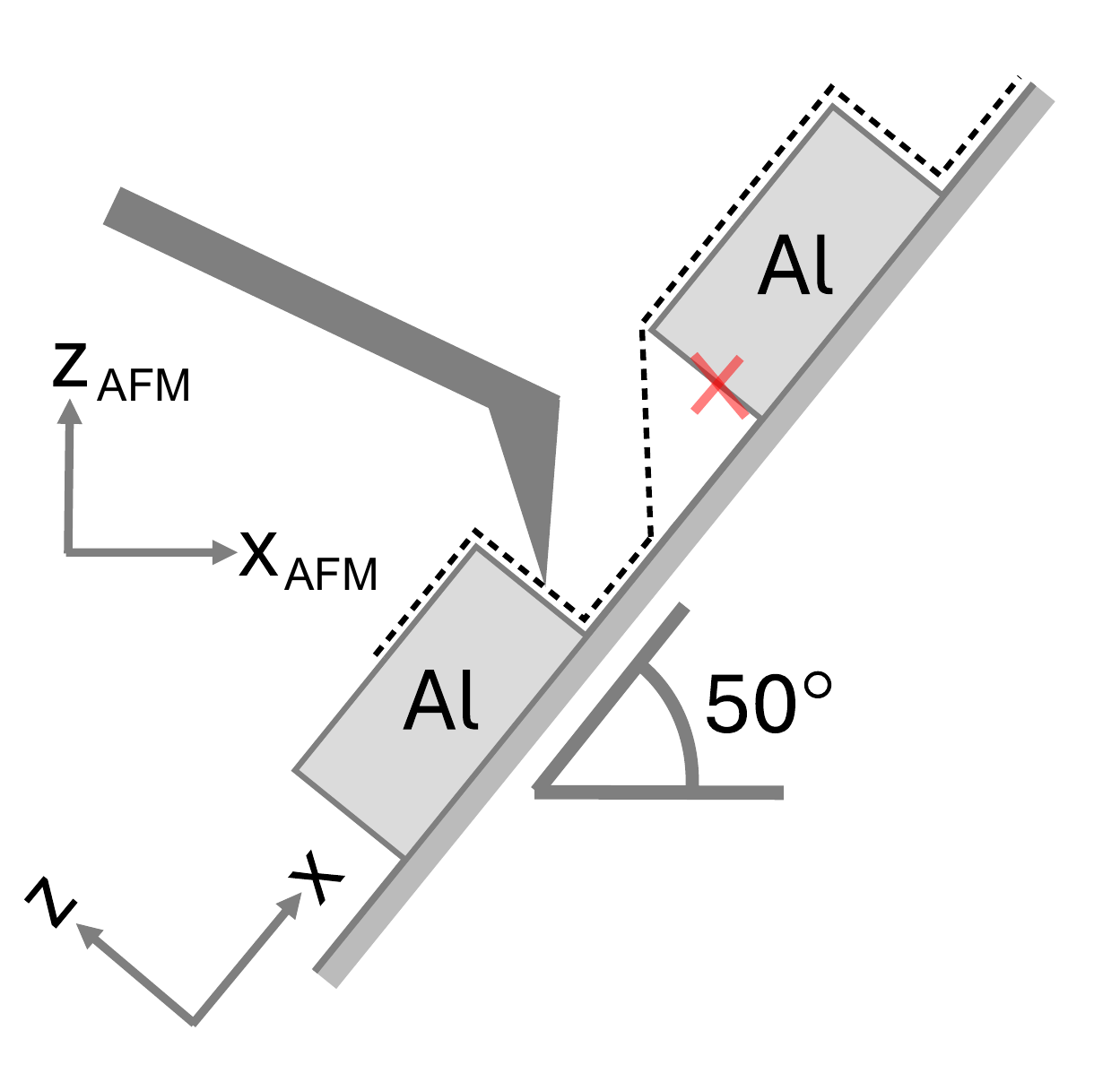}
    \caption{Schematic of the tilted-sample AFM approach enabling quantitative analysis of sidewall geometries, including undercut structures. The red crosses indicate vertical sidewall regions that are not accessible and cannot be reconstructed using the conventional non-tilted approach. The sampled data of a sidewall profile are acquired in a rotated coordinate system $(x_{\mathrm{AFM}}, z_{\mathrm{AFM}})$ relative to the substrate plane.}
	\label{fig:SketchAPP}
\end{figure}

The Lateral Sidewall Profile (LSP), $X_{\mathrm{LSP},i}(y)$, see Fig.~\ref{fig:Sketch}, describes the lateral position of the sidewall projected onto a plane parallel to the substrate, where $y$ denotes the coordinate along the nominal sidewall direction. The lateral sidewall profile $R_q^{\mathrm{LSP}}$ is obtained by projecting the measured AFM surface from the tilted measurement coordinate system onto the substrate plane. The index $i$ identifies the individual profiles in the set of measured profiles and corresponds to their vertical
position along the sidewall, from the foot to the top of the sidewall. For the measurement geometry used in this work, this projection is approximated by 
\begin{equation}
X_{\mathrm{LSP},i}(y)
\approx
Z_{\mathrm{AFM},i}(x_{\mathrm{AFM}},y)\cos\theta ,
\label{eq:X_LSP_projection}
\end{equation}
where $\theta$ is the angle between the substrate normal and the AFM measurement direction, as illustrated in Fig.~\ref{fig:SketchAPP}. 

The amplitude-based lateral sidewall roughness are calculated relative to the designed target geometry, $X_{\mathrm{ref}}(y)$, which may be straight or curved. For a straight target geometry, a least-squares linear reference profile is subtracted from each measured $X_{\mathrm{LSP}, i(y(}$, yielding the residual profile
\begin{equation}
\Delta X_{\mathrm{LSP},i}(y) = X_{\mathrm{LSP},i}(y)-X_{\mathrm{ref}}(y),
\label{eq:DeltaX_LSP}
\end{equation}
which represents the fabrication-induced deviations from the ideal sidewall geometry.

Before calculating the sidewall descriptors, the residual profiles $\Delta X_{\mathrm{LSP},i}(y)$ are filtered using high- and low-pass filters to remove measurement artefacts and spatial frequencies outside the defined measurement bandwidth. In analogy with conventional surface texture measurements, the spatial bandwidth is defined by an S-filter cutoff wavelength, $\lambda_S$, and an L-filter cutoff wavelength, $\lambda_C$, where $\lambda_S$ defines the short-wavelength limit and $\lambda_C$ defines the long-wavelength limit of the measured profile. The selected cutoff wavelengths are determined by the physically relevant length scales of the sidewall geometry. The resulting filtered residual profiles, $\Delta X_{\mathrm{LSP},i}^{\,f}(y)$, are used for calculation of the sidewall descriptors.

The Gaussian filter width is selected to provide 50\% amplitude transmission at the corresponding cutoff wavelength:
\begin{equation}
\sigma = 0.1874\lambda ,
\label{eq:GaussianFilterWidth}
\end{equation}
where $\lambda$ corresponds to either the short-wavelength cutoff $\lambda_S$ or the long-wavelength cutoff $\lambda_C$. Further details are given in Appendix~\ref{sec:RelationBetweenISOGaussian}.

The RMS roughness, $R_q$, is selected as a descriptor of the residual lateral sidewall amplitude variation because it provides a well-established statistical measure of profile amplitude. The exponent of two in the RMS definition emphasizes larger deviations while providing a simple and widely used metric for comparison of surface amplitude variations.

The individual RMS lateral sidewall roughness values, $R_{q,i}^{\mathrm{LSP}}$, are calculated from the filtered residual lateral sidewall profiles, $\Delta X_{\mathrm{LSP},i}^{\,f}(y)$:
\begin{equation}
R_{q,i}^{\mathrm{LSP}} = \sqrt{
\frac{1}{L}
\int_0^L
\left(
\Delta X_{\mathrm{LSP},i}^{\,f}(y)
\right)^2
\,\mathrm{d}y }.
\label{eq:Rq_LSP_i}
\end{equation}
The reported RMS lateral sidewall roughness,
$R_q^{\mathrm{LSP}}$, is defined as the average RMS amplitude of the measured profiles across the sidewall:
\begin{equation}
R_q^{\mathrm{LSP}} = \frac{1}{N}
\sum_{i=1}^{N}
R_{q,i}^{\mathrm{LSP}} .
\label{eq:Rq_LSP}
\end{equation}
where $N$ is the number of measured LSP profiles along the vertical extent of
the sidewall and $L$ is the length of each individual sidewall profile.

\subsection{The vertical sidewall profile} 
Since no target geometry is available for the Vertical Sidewall Profile (VSP), see Fig.~\ref{fig:Sketch}, a residual amplitude descriptor such as $R_q$ cannot be meaningfully defined. Instead, descriptors based on the local radius of curvature provide a measure of the interface geometry, quantifying regions with small radii of curvature that are expected to concentrate electric fields and thereby influence energy dissipation.

The vertical sidewall is characterized by a set of individual measured profiles $Z_{\mathrm{VSP},i}(x) = Z_{\mathrm{AFM},i}(x_{\mathrm{AFM}})$, where $i$ denotes the position of the profile along the sidewall in the nominal $y$-direction. Each profile represents the vertical sidewall surface in a cross-sectional plane perpendicular to the nominal sidewall direction, with $x$ representing the coordinate normal to the sidewall. Prior to calculating the curvature, each individual profile is filtered using a Gaussian S-filter with a short-wavelength cutoff of $\lambda_S$ to suppress measurement noise and spatial wavelengths that cannot be reliably resolved due to the finite AFM probe geometry. The resulting filtered individual profile is denoted $Z_{\mathrm{VSP},i}^{f}(x)$.

The local observed curvature $\kappa_i(x)$ is calculated from the filtered vertical sidewall profile $Z_{\mathrm{VSP},i}^{f}(x)$. For a profile expressed in Cartesian coordinates, the local curvature is defined as
\begin{equation}
\kappa_i(x)=
\frac{ {Z_{\mathrm{VSP},i}^{f}}''(x) }{ \left(1+\left({Z_{\mathrm{VSP},i}^{f}}'(x)\right)^2\right)^{3/2} },
\label{eq:LocalCurvature}
\end{equation}
where $Z_{\mathrm{VSP},i}^{f}(x)$ is the filtered individual vertical sidewall profile expressed in the observed AFM coordinate system. 

For surface features where both the probe apex and the sample feature can locally be approximated by circular arcs, tip--sample convolution can be approximated by an additive relationship between the observed radius of curvature, the effective probe radius, and the sample radius of curvature:
\begin{equation}
\rho_{\mathrm{OBS}}(x) \approx \rho(x) - \rho_{\mathrm{probe}}^{\mathrm{eff}}.
\end{equation}
Here, $\rho_{\mathrm{OBS}}(x)$ is the signed radius of curvature obtained from the AFM measurement, $\rho_{\mathrm{probe}}^{\mathrm{eff}}$ is the effective probe radius and is defined as a positive quantity, see Sec~\ref{sec:Curvature-BasedRadiusEstimation}, and $\rho(x)$ is the signed radius of curvature of the sample feature. With the coordinate convention used here, $\rho(x)$ is negative for surface protrusions and positive for depressions or holes. The approximation can be applied to both convex protruding and concave features, provided that the local geometry is accessible to the probe and that the curvature contribution from the probe can be represented by an effective radius. Limitations associated with hypothetical extremely sharp concave features that cannot be accessed by the probe are discussed below.

The local radius of curvature corrected for the probe radius is therefore given by
\begin{equation}
\rho(x) \approx \rho_{\mathrm{OBS}}(x) + \rho_{\mathrm{probe}}^{\mathrm{eff}} = \frac{1}{\kappa_{\mathrm{OBS}}(x)} + \rho_{\mathrm{probe}}^{\mathrm{eff}},
\label{eq:RadiusOfCurvature}
\end{equation}
where $\kappa_{\mathrm{OBS}}(x)$ is the signed local curvature calculated from the measured AFM profile. The corresponding probe-radius-corrected local curvature is
\begin{equation}
\kappa(x) = \frac{1}{\rho(x)} = \frac{\kappa_{\mathrm{OBS}}(x)}
{1+\rho_{\mathrm{probe}}^{\mathrm{eff}}\kappa_{\mathrm{OBS}}(x)}.
\label{eq:CorrectedCurvature}
\end{equation}
 To avoid reporting protrusion radii with magnitudes smaller than the effective probe radius, corrected radii satisfying $\rho(x)\ge -\rho_{\mathrm{probe}}^{\mathrm{eff}}$ are assigned the limiting value $-\rho_{\mathrm{probe}}^{\mathrm{eff}}$. The corresponding probe-radius-corrected local curvature is subsequently obtained as $\kappa_i(x)=1/\rho_i(x)$. An RMS-equivalent mean curvature, $\kappa_{\mathrm{RMS},i}$, with units of inverse length, is defined for each individual profile as the root-mean-square value of the local curvature along the profile. To account for the actual geometry of the curved profile, the averaging is performed along the arc length:
\begin{equation}
\kappa_{\mathrm{RMS},i} = \left(
\frac{
\int_{x_1}^{x_2}
\kappa_i(x)^2
\left[1+\left({Z_{\mathrm{VSP},i}^{f}}'(x)\right)^2\right]^{1/2}\,\mathrm{d}x}{
\int_{x_1}^{x_2}
\left[1+\left({Z_{\mathrm{VSP},i}^{f}}'(x)\right)^2\right]^{1/2}\,\mathrm{d}x}
\right)^{1/2}.
\label{eq:kappaRMSintegral}
\end{equation}
The profile-averaged RMS curvature is defined as the arithmetic mean of the values obtained from the $N$ individually evaluated vertical sidewall profiles:
\begin{equation}
\overline{\kappa}_{\mathrm{RMS}} = \frac{1}{N}
\sum_{i=1}^{N}
\kappa_{\mathrm{RMS},i}.
\label{eq:kappaRMSaverage}
\end{equation}
The curvature-equivalent radius, $R_{\rho}$, is defined as the inverse of the profile-averaged RMS curvature:
\begin{equation}
R_{\rho}
=
\frac{1}{\overline{\kappa}_{\mathrm{RMS}}}.
\label{eq:Rkappa}
\end{equation}
Small values of $R_{\rho}$ indicate the presence of many sharp features, where the surface profile locally matches the curvature of associated small circles. Small local radii of curvature contribute to $\kappa_{\mathrm{RMS}}$ and therefore reduce $R_\rho$, independent of whether they correspond to protruding or recessed surface features.
\subsection{The minimum radius of curvature}
The local radius of curvature, $\rho(x)$, can take both positive and negative values. For a profile described by $z(x)$, the local radius of curvature is given by
\begin{equation}
\rho(x)=
\frac{\left(1+\left(\frac{\mathrm{d}z}{\mathrm{d}x}\right)^2\right)^{3/2}}
{\frac{\mathrm{d}^2z}{\mathrm{d}x^2}}.
\end{equation}
With the coordinate convention used here, negative values of $\rho(x)$ correspond to \textbf{protrusions}, whereas positive values correspond to \textbf{depressions or holes}. The minimum radius of curvature is therefore defined separately for the two types of features. For protrusions, the descriptor is the negative radius with the smallest absolute value,
\begin{equation}
\rho_{\min}^{\mathrm{prot}} = -\min_{\rho(x)<0} \lvert \rho(x) \rvert,
\label{eq:rhomin1}
\end{equation}
whereas for depressions or holes it is defined as
\begin{equation}
\rho_{\min}^{\mathrm{dep}} = \min_{\rho(x)>0} \rho(x).
\label{eq:rhomin2}
\end{equation}
Thus, $\rho_{\min}^{\mathrm{prot}}$ remains negative, while $\rho_{\min}^{\mathrm{dep}}$ is positive. The $\rho_{\min}$ descriptor serves two purposes. When the measurement quality is sufficient to ensure reliable curvature determination, $\rho_{\min}$ may be interpreted as an estimate of the smallest local radius of curvature present on the sidewall, thereby identifying the sharpest resolvable geometrical feature of the measured profile. Furthermore, $\rho_{\min}$ is used to determine the Curvature-Based Probe Radius Estimation, see Appendix~\ref{sec:Curvature-BasedRadiusEstimation}.

%
%%
%%%
\section{Curvature-Based Probe Radius Estimation (CPRE)}
\label{sec:Curvature-BasedRadiusEstimation}
%%%
%%
%
Blind probe reconstruction can estimate the probe geometry from measured surface topography, but for the probe shapes within few nanometers from the apex, which is the relevant length scale for this work, the reconstruction depends critical on the treshold values for the auxiliary parameter to suppress noise in the image used to reconstruct the probe shape~\cite{Villarrubia97, TianEA08, ErinEA14}. Sharp features~\cite{RaoThesis} and sharp grating profiles~\cite{BoonpuekFelts2023} have also been used to estimate an effective AFM probe radii. 

In the present work, an approach based on the local curvature of multiple sharp protrusions is used to obtain a conservative upper bound, referred to as Curvature-Based Probe Radius Estimation (CPRE). Rather than reconstructing the complete probe geometry, CPRE estimates an effective probe radius from measurements on a tip characterization sample and uses this estimate to correct the measured local radii of curvature. For an ideal round probe the estimated effective probe radius would be equal to the radius of the round probe. The method provides a practical probe radius correction together with an associated measurement uncertainty suitable for quantitative AFM measurements.

\subsection{Theory} 
\label{TheoryCPRE-Approach}
In the CPRE approach, a dedicated tip-characterization sample containing sharp features is measured using the AFM probe under validation. For a single averaged measured profile, six local minima of the radius of curvature associated with protrusions are identified. The radius of curvature at the $j$-th local minimum is denoted by
\begin{equation}
\rho_{\mathrm{obs},j}^{\mathrm{prot}}
=
\rho_{\mathrm{obs}}(x_j),
\qquad
\rho_{\mathrm{obs}}(x_j)<0,
\qquad
j=1,\ldots,6,
\end{equation}
where $x_j$ denotes the position of the $j$-th local minimum. The upper-bound estimate for the effective probe radius is then determined from the average absolute radius of the $m = 6$ local minima,
\begin{equation}
\label{RhoUB6Prot}
\rho_{\mathrm{probe}}^{\mathrm{UB}}
=
\frac{1}{m}
\sum_{j=1}^{m}
\lvert
\rho_{\mathrm{obs},j}^{\mathrm{prot}}
\rvert.
\end{equation}
The upper-bound estimate for the effective probe radius is based on the observed radius of curvature of protrusions, $\rho_{\mathrm{obs}}^{\mathrm{prot}}$. The corresponding radius of curvature for depressions, $\rho_{\mathrm{obs}}^{\mathrm{dep}}$, is not used because it cannot provide an upper bound for the effective probe radius.

Assuming that the minimum radius of curvature of the characterization sample and the effective probe radius are of comparable magnitude, the effective probe radius can be estimated as
\begin{equation}
\rho_{\mathrm{probe}}^{\mathrm{eff}} = \frac{1}{2}\rho_{\mathrm{probe}}^{\mathrm{UB}} .
\label{RhoEffEqualHalfOBS}
\end{equation}

The measured local radius of curvature is subsequently corrected for the probe contribution according to
\begin{equation}
\label{CorrectedForTheProbe}
\rho_i(x)= \begin{cases} \rho_{\mathrm{obs},i}(x)+\rho_{\mathrm{probe}}^{\mathrm{eff}} &\mathrm{if}\hspace{8pt} \rho_{\mathrm{obs},i}(x)+\rho_{\mathrm{probe}}^{\mathrm{eff}} \le -\rho_{\mathrm{probe}}^{\mathrm{eff}}, \\[6pt] -\rho_{\mathrm{probe}}^{\mathrm{eff}} &\mathrm{if}\hspace{8pt} -\rho_{\mathrm{probe}}^{\mathrm{eff}} < \rho_{\mathrm{obs},i}(x)+\rho_{\mathrm{probe}}^{\mathrm{eff}} < 0, \\[6pt] \rho_{\mathrm{obs},i}(x)+\rho_{\mathrm{probe}}^{\mathrm{eff}} &\mathrm{if}\hspace{8pt} \rho_{\mathrm{obs},i}(x)+\rho_{\mathrm{probe}}^{\mathrm{eff}} \ge 0. \end{cases}
\end{equation}
where $\rho_i(x)$ represents the probe-corrected local radius of curvature of profile $i$. To avoid reporting protrusion radii with magnitudes smaller than the effective probe radius, corrected values for which $\rho_i(x)\ge -\rho_{\mathrm{probe}}^{\mathrm{eff}}$ are assigned the limiting value $-\rho_{\mathrm{probe}}^{\mathrm{eff}}$.

\subsection{Uncertainty} 
The uncertainty associated with the limited knowledge of the effective probe radius is evaluated conservatively by assuming a rectangular distribution over the interval from zero to the upper bound of the probe radius of curvature, $\rho_{\mathrm{probe}}^{\mathrm{UB}}$:
\begin{equation}
\label{uRhoEffProbe}
u(\rho_{\mathrm{probe}}^{\mathrm{eff}}) = \frac{\rho_{\mathrm{probe}}^{\mathrm{UB}}}{2\sqrt{3}},
\end{equation}
where $2\sqrt{3}$ is the standard deviation factor for a rectangular distribution. 

\subsection{Traceability}
The traceability of the correction term $\rho_{\mathrm{probe}}^{\mathrm{eff}}$ is established through the calibration of the AFM measurement system. The traceability of the CPRE approach does not require the characterization sample to be a certified reference material, since the correction term is derived from the calibrated AFM measurement and the corresponding probe-characterization procedure.

CPRE treats the unknown contribution from the finite radius of curvature of the characterization sample as part of the uncertainty evaluation. Consequently, the method remains traceable through the AFM calibration, while the uncertainty reflects the limited knowledge of the actual sharpness of the characterization sample. A characterization sample with reduced sharpness will result in a larger uncertainty; however, the estimated probe radius remains valid within the associated uncertainty limits.

\subsection{Limitation} 
\label{Limitation}
The simple correction model $\rho_i(x) = \rho_{\mathrm{obs},i}(x) - \rho_{\mathrm{probe}}^{\mathrm{eff}}$ represents a local curvature correction based on an effective probe radius. It does not attempt to reconstruct inaccessible geometric information. In particular, hypothetical extremely sharp concave features that cannot be accessed by the probe are not assessed, as no assumed detailed knowledge of the probe shape can recover information that is absent from the measured AFM profile. This limitation is analogous to conventional surface roughness measurements, where probe convolution effects could alos be present but are not normally considered. Because the correction term $\rho_{\mathrm{probe}}^{\mathrm{eff}}$ has an 
associated standard uncertainty $u(\rho_{\mathrm{probe}}^{\mathrm{eff}})$, corrected radii of curvature smaller than $\rho_{\mathrm{probe}}^{\mathrm{UB}}$ may be numerically obtained. However, such estimates cannot be distinguished from the probe-radius upper bound within the uncertainty introduced by the probe-radius correction.

\section{AFM measurement validity and acceptance criteria}
\label{sec:ValidityAndAcceptanceCriteria}

Amplitude-modulation AFM can exhibit apparent height shifts (often in the order of less than a nm) and other imaging artifacts~\cite{Kyhle1998,Eves2012,Temiryazev2021} associated with the nonlinear tip-sample interaction and transitions between attractive- and repulsive-dominated oscillation states, which can produce changes in amplitude and phase. Such dynamical effects may influence the measured profile and hence the derived radius of curvature, particularly near regions with small radius of curvature. Therefore, all measurements were checked for signs of instability and related artifacts~\cite{Kyhle1998,Eves2012,Temiryazev2021}. As acceptance criteria for measurement validity the following were used: (i) no distinct jump in the phase along the profile, in particular near regions with small radius of curvature, (ii) the phase variation was required to be smooth and remain within $\pm 5^\circ$, (iii) the difference between trace and retrace profiles was required to remain within the acceptable measurement variation, and (iv) the derived radius of curvature was required to be insensitive to reasonable changes in drive frequency and amplitude setpoint.

\section{Experimental estimation of the effective probe radius}
\label{sec:ExperimantalCPRE}
\begin{figure}[b!]
    \centering
    \includegraphics[width=0.95\linewidth]{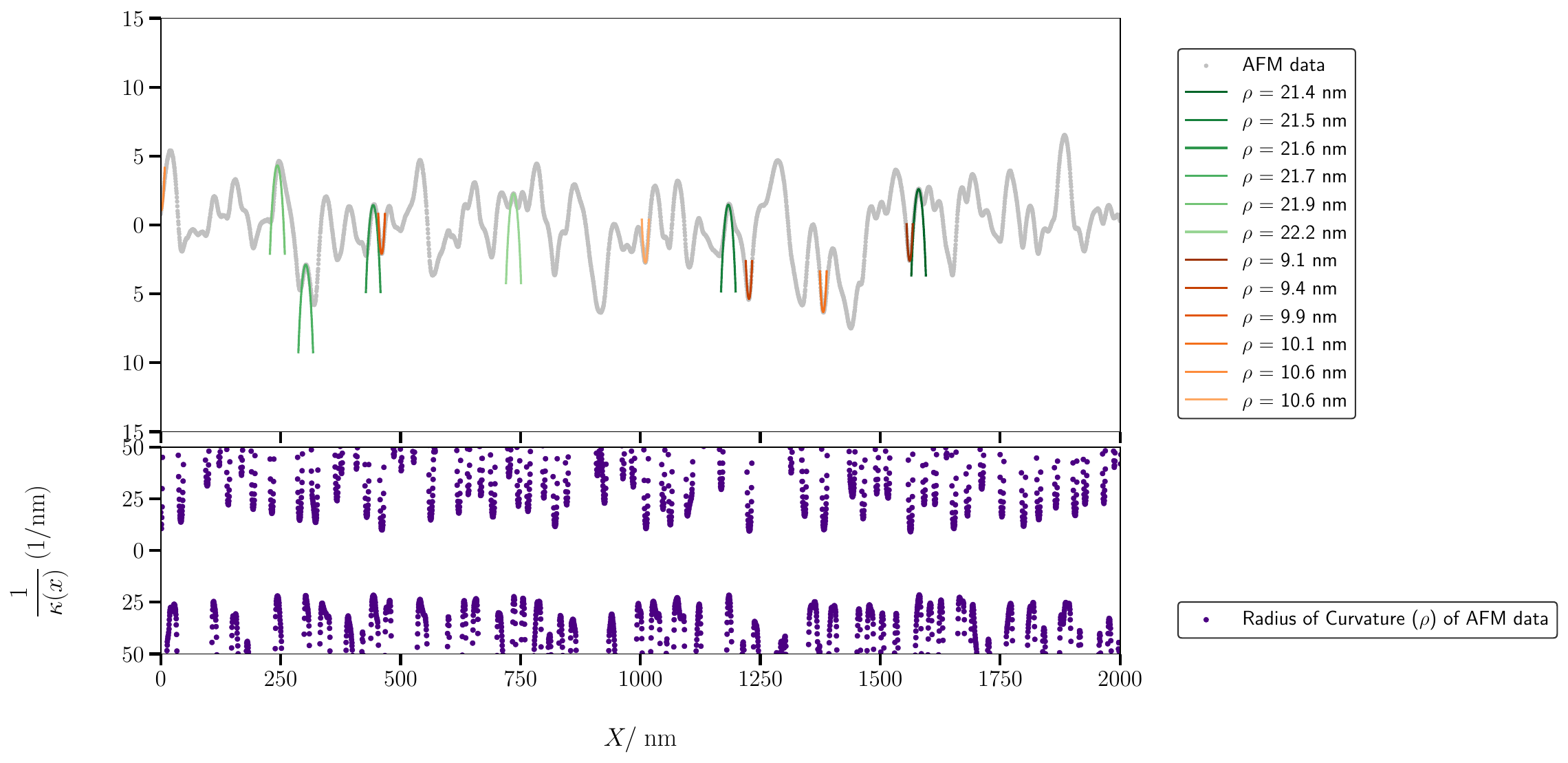}
    \caption{Profile of the tip characterization sample recorded by AFM. The top panel shows a profile averaged over 93 individual profiles recorded with a used probe. The green elongated circular arcs indicate the six protrusions with the smallest radii of curvature, and the green elongated arcs indicate the six depressions with the smallest radii of curvature. The lower panel shows the local observed radius of curvature $1/\kappa_{\mathrm{OBS}}(x)=\rho_{\mathrm{OBS}}(x)$. The smooth variation of the local observed radius of curvature indicates consistent and reliable data.}
    \label{fig:Profile_Nioprobe}
\end{figure}

\begin{figure}[t!]
    \centering
    \includegraphics[width=0.6\linewidth]{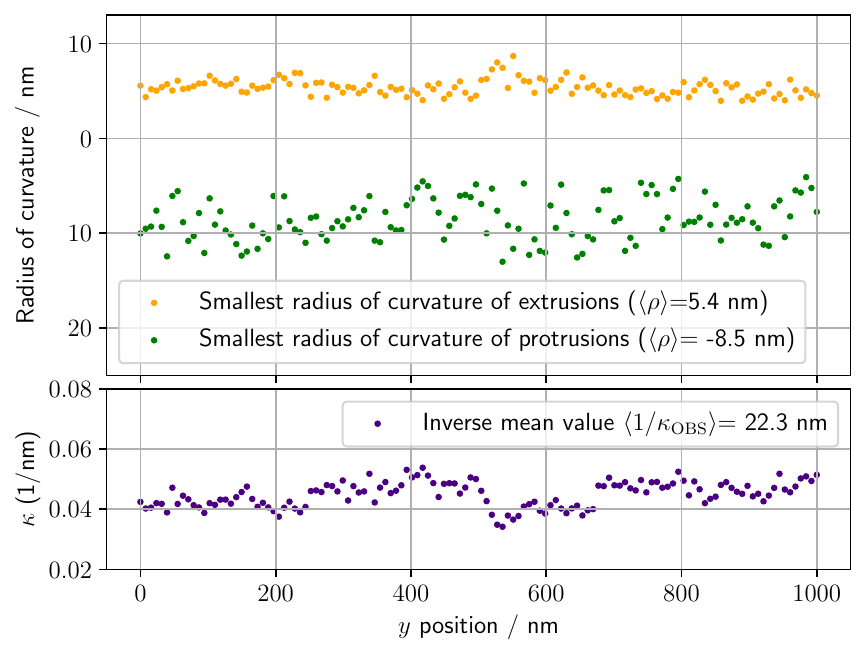}
    \caption{Analysis of an ensemble of profiles of a tilted tip characterization sample. The upper panel shows the smallest observed radii of curvature of the protrusions, $\rho_{\mathrm{obs}}^{\mathrm{prot}}$, and depressions, $\rho_{\mathrm{obs}}^{\mathrm{dep}}$, for each of the 128 individual profiles as a function of the profile position along the $y$-direction. The lower panel shows the value of $\kappa_{\mathrm{OBS}}$, calculated from the observed profile according to Eq.~\eqref{eq:kappaRMSintegral}, for each profile position along the $y$-direction.}
    \label{fig:Kappa_NioProbe}
\end{figure}

The tip characterization sample used consisted of a thin film of nanoscale sharp protrusions~\cite{NioProbe}, reported to be niobium with protrusion radii less than 5 nm~\cite{Liu2008}. The sample was imaged under the same measurement conditions as for the sidewall. A conservative upper limit for the observed $\rho_{\mathrm{obs}}$ was estimated based on a profile averaged over 93 individual profiles of a used probe, where the average radius of the six smallest observed radius of curvature of the averaged profile was found to be 21.7(0.3) $\mathrm{nm}$ calculated according to Eq.~\eqref{RhoUB6Prot}, see Fig.~\ref{fig:Profile_Nioprobe}. Averaging 93 profiles reduces random noise and suppresses short-term drift, making the estimated average a conservative upper limit. Based on the above measurement and other measurements for different probes, the upper bound $\rho_{\mathrm{probe}}^{\mathrm{UB}}$ for the effective probe radius for the batch of probes used was conservatively estimated to be 21.7 nm. This gives an effective probe radius for the batch of probe used according to Eq.~\eqref{RhoEffEqualHalfOBS} of
\begin{equation}
\rho_{\mathrm{probe}}^{\mathrm{eff}} = 10.9\,\mathrm{nm}.
\end{equation}
The associated uncertainty is conservatively estimated according to Eq.~\eqref{uRhoEffProbe} to be $u(\rho_{\mathrm{probe}}^{\mathrm{eff}}) = 6.3\,\mathrm{nm}$. The observed average minimum radius of curvature on each of $n = 128$ tilted profiles in an image (see Fig.~\ref{fig:Kappa_NioProbe}) was found to be
\begin{equation}
\overline{\rho}_{\mathrm{obs}}^{\mathrm{prot}}
=
\frac{1}{n}
\sum_{i=1}^{n}
\lvert
\rho_{\mathrm{obs},i}^{\mathrm{prot}}
\rvert
=
8.5 \, \mathrm{nm}.
\label{eq:RhoObsProtAverage}
\end{equation}
This confirms that the upper bound $\rho_{\mathrm{probe}}^{\mathrm{UB}} \, = 21.7 \, \mathrm{nm} \, > \overline{\rho}_{\mathrm{obs}}^{\mathrm{prot}}$ is a conservative choice, allowing small variations of probe shape, wear and the possible influence of drift and noise on each profile with a good margin.

The experimentally estimated effective radius is larger than the manufacturer's specification of $\rho < 10$ nm. This is consistent with the possibility of probe wear during use, which may increase the effective radius of curvature, and with the conservative approach outlined here. 

\subsection{Treatment of sharp protrusions} 
For very sharp protrusions, the probe-radius correction in Eq. ~\eqref{eq:CorrectedCurvature} can become unreliable if $1+\rho_{\mathrm{probe}}^{\mathrm{eff}}\kappa_{\mathrm{OBS}}$ approaches zero, i.e., if $\kappa_{\mathrm{OBS}}$ approaches $-1/\rho_{\mathrm{probe}}^{\mathrm{eff}}$. To avoid this unreliable regime while maintaining the contribution of sharp protrusions to the curvature descriptor, a conservative lower limit of $21.8~\mathrm{nm}$ is applied on the observed radius of protrusions. Thus, if the observed protrusion radii satisfy $-21.8~\mathrm{nm} < \rho_{\mathrm{OBS},i}(x) <0$, it is assigned the limit value $\rho_{\mathrm{OBS},i}(x) = -21.8~\mathrm{nm}$
corresponding to $\kappa_{\mathrm{OBS},i}(x) =-0.0459~\mathrm{nm}^{-1}$. With $\rho_{\mathrm{probe}}^{\mathrm{eff}}=10.9~\mathrm{nm}$, the resulting corrected curvature is limited to $\kappa_i(x)=-0.0917~\mathrm{nm}^{-1}$. Positive observed radii, corresponding to depressions, are not affected by this limiting procedure. The limit is conservative; several radii of around $21.8~\mathrm{nm}$ and below have been reliably measured on the tip characterization sample. In the hypothetical case that all observed protrusion radii were smaller than $21.8~\mathrm{nm}$, the corrected radius would still be limited to $-10.9~\mathrm{nm}$. Taking into account the associated measurement uncertainty $u(\rho_{\mathrm{probe}}^{\mathrm{eff}})=6.3~\mathrm{nm}$, a radius significantly smaller than $21.8~\mathrm{nm}$ cannot be measured.

%
%%
%%%
\section{Sources of measurement variability}
\label{SourcesMeasurementVariability}
%%%
%%
%
%
%%
\subsection{Probe-to-probe variation}
An important source of variability in the sidewall descriptors is probe-to-probe variation; i.e., different probes may yield different descriptor values for the same sidewall geometry. Although probes may have the same nominal tip radius (e.g., $\rho_{\mathrm{probe}} < 10$~nm), variations in the fabrication process result in differences in the actual probe geometry. The standard uncertainty associated with the above variation is estimated to be 10--20\% for $R^{\mathrm{LSP}}_q$. For the probe used in this work, the effective probe radius is assigned a conservatively estimated standard uncertainty of $u(\rho_{\mathrm{probe}}^{\mathrm{eff}})=6.3$~nm, (see App.~Sec.~\ref{sec:ExperimantalCPRE}) which contributes to the uncertainty of $R_\rho$ through the proberadius correction given in Eq.~\ref{eq:LocalCurvature} to~\ref{eq:Rkappa}. In general the standard uncertainty associated with variation in the probe radius is estimated to be 5 - 8 nm. 

\subsection{Probe degradation and instability}
Another source of variability is probe degradation and contamination. Probe wear or material transfer from the sample surface during scanning may alter the effective probe geometry and measurement response. Based on experience from numerous measurements, most probes are suitable for approximately one day of measurements. The standard uncertainty associated with the above variation is estimated to be 5--15\% for both $R^{\mathrm{LSP}}_q$ and $R_\rho$.

\subsection{Spatial inhomogeneity of the sample}
The reported descriptors represent averages over the investigated microscopic regions ($<$100~$\mu$m in lateral size). The descriptor values may vary between different microscopic regions due to local inhomogeneity in the fabrication process. This larger-scale variation is not captured by a single measurement and is therefore not included in the reported descriptor values. It should be evaluated separately and it should be stated separately for what area(s) the measurement are valid. The variation associated with with sample inhomogeneity for the superconducting elements investigated has been in the range of 5--15\% for both $R^{\mathrm{LSP}}_q$ and $R_\rho$.

\subsection{Particle and contamination}
The descriptors can also be influenced by particles and other surface contamination originating from the fabrication process or the measurement environment. Such features may significantly affect device functionality, and their presence should be documented and quantified. However, particles may consist of different materials introduced after fabrication, and it remains unclear whether they should be included in descriptors intended to describe the designed geometry of the sidewall. Furthermore, particles and contamination cannot be meaningfully described by the sidewall descriptors focused on average amplitude and surface curvature. The recommended approach is therefore to minimize or remove the influence of particles and contamination when determining the sidewall descriptors defined in this work. Ideally, new descriptors should be defined and recorded to quantify the presence and amount of particles and contamination; however, this is outside the scope of the present work. The standard uncertainty associated with the above variation is estimated to be up to 50 \% for both $R^{\mathrm{LSP}}_q$ and $R_\rho$ before the measurement is classified as an outlier.

\subsection{Additional experiment parameters}
General measurement conditions reported in the literature, but not specifically addressed in this paper, may contribute to the observed variation in the descriptors. These include the applied probe force, thermal stability, environmental vibrations, and, importantly, the adjustment of the feedback loop, which effectively acts as an additional filtering process. The standard uncertainty associated with the above variation is estimated to be up to 20\% for both $R^{\mathrm{LSP}}_q$ and $R_\rho$. 

\subsection{Data processing}
General aspects of data processing reported in the literature, but not specifically addressed in this paper, may also contribute to variation in the descriptors, in particular the treatment of outliers, the exact numerical implementation of the calculations, and the application of filters, including, e.g., the order in which high- and low-pass filtering is performed. The standard uncertainty associated with the above variation is estimated to be up to 10\% for both $R^{\mathrm{LSP}}_q$ and $R_\rho$. For a summary of sources of variability, see Table~\ref{tab:variability}.

%
%%
%%%
\section{Improving measurement repeatability}
\label{Improving-Masurement-Repeatability}
%%%
%%
%
This subsection describes procedures for improving repeatability and enabling accurate comparison of sidewall descriptors. Using the same probe, short-term changes in probe condition (timescales of less than one day) can be monitored and mitigated by performing reference measurements on a reference sample before and after a series of measurements. If the reference measurements remain within predefined limits, the measurements performed between the reference measurements can be considered to have a low relative variation.

Using clean surfaces without particles or contamination is the preferred approach to minimize variation from this source. In this work it is documented that this is possible, and it is estimated that the influence of particles and contamination on the images shown is close to zero \%. Alternatively, suspected particles and contamination can be identified and removed during image processing, for example by excluding affected regions. With this approach, the residual variation can also be limited to approximately 1-2 \% (not shown). Further improvement in repeatability can be achieved by maintaining identical measurement conditions, including applied force, feedback parameters, and other instrument settings. Finally, consistent data processing using identical algorithms and processing sequences can reduce additional numerical variation.

Fig.~\ref{fig:Correlations}(h) shows a total of 10 measurements performed with three different probes on four different samples. The average observed standard deviation, $s(R_q^{\mathrm{LSP}})=4.1\%$, obtained using the mitigation strategies described above, represents a conservative upper limit for the contribution from each individual source of variation, including probe-to-probe variation, probe degradation, experimental parameters, and data processing.

%
%%
%%%
\section{Relation between ISO Gaussian filtering and standard deviation formulation}
\label{sec:RelationBetweenISOGaussian}
%%%
%%
%
The Gaussian S- and L-filters used in this work follow the ISO~16610-21
(formerly ISO~11562) definition, where the filter is specified by its cutoff wavelength, $\lambda$. The normalized spatial weighting function is
\begin{equation}
G(x)=
\frac{1}{\alpha\lambda}
\exp\left[
-\pi
\left(
\frac{x}{\alpha\lambda}
\right)^2
\right],
\end{equation}
where $\alpha=0.4697$. This definition corresponds to a Gaussian filter with
50\% amplitude transmission at the cutoff wavelength.

The Gaussian filter can equivalently be expressed in terms of its standard
deviation,
\begin{equation}
G(x)=
\frac{1}{\sigma\sqrt{2\pi}}
\exp\left(
-\frac{x^2}{2\sigma^2}
\right).
\end{equation}
Equating the exponents of the two Gaussian expressions gives
\begin{equation}
\sigma =\frac{\alpha\lambda}{\sqrt{2\pi}} \approx 0.1874\,\lambda.
\end{equation}
Thus, the ISO Gaussian filter is mathematically equivalent to the standard Gaussian parameterization, with the filter width expressed by the standard deviation rather than the cutoff wavelength.

\end{appendix}
\end{document}